\documentclass[10pt,letterpaper]{article}
\usepackage[top=0.85in,left=1in,footskip=0.75in]{geometry}

\usepackage{amsmath,amssymb}

\usepackage{changepage}

\usepackage{textcomp,marvosym}

\usepackage{cite}

\usepackage{nameref,hyperref}

\usepackage[right]{lineno}

\usepackage[nopatch=eqnum]{microtype}
\DisableLigatures[f]{encoding = *, family = * }

\usepackage[table]{xcolor}

\usepackage{array}

\usepackage[aboveskip=1pt,labelfont=bf,labelsep=period,justification=raggedright,singlelinecheck=off]{caption}

\makeatletter
\renewcommand{\@biblabel}[1]{\quad#1.}
\makeatother

\usepackage{graphicx}
\usepackage{booktabs}
\usepackage{caption,subcaption}
\usepackage{float}
\usepackage{amsmath}
\usepackage{xr-hyper}
\usepackage{hyperref}
\usepackage{multirow}
\usepackage{authblk}

\newcommand{\siref}[1]{SI \ref{#1}}

\newcommand{\suppfig}[1]{%
    \captionsetup{labelformat=empty}
  \caption{\textbf{Figure S.\thefigure}: #1}%
  \label{fig:supp:#1}%
}

\newcommand{\supptable}[1]{%
\captionsetup{labelformat=empty}
  \caption{\textbf{Table S.\thetable}:#1}%
  \label{tab:supp:#1}%
}

\begin{document}

\vspace*{0.2in}

\begin{flushleft}
{\Large
\textbf\newline{Socioeconomic determinants of protective behaviors and contact patterns in the post-COVID-19 pandemic era: a cross-sectional study in Italy} 
}
\newline

Michele Tizzani\textsuperscript{1},
Laetitia Gauvin\textsuperscript{2,1}
\\
\bigskip
\textbf{1} ISI Foundation, Turin, Italy
\\
\textbf{2} UMR 215 PRODIG, Institute for Research on Sustainable Development - IRD, 5 cours des Humanités, F-93 322 Aubervilliers Cedex, France
\\

\bigskip

\end{flushleft}

\begin{abstract}

Socioeconomic inequalities significantly influence infectious disease outcomes, as seen with COVID-19, but the pathways through which socioeconomic conditions affect transmission dynamics remain unclear. To address this, we conducted a survey representative of the Italian population, stratified by age, gender, geographical area, city size, employment status, and education level. The survey's final aim was to estimate differences in contact and protective behaviors across various population strata, both being key components of transmission dynamics.
Our initial insights based on the survey indicate that years after the pandemic began, the perceived impact of COVID-19 on professional, economic, social, and psychological dimensions varied across socioeconomic strata, extending beyond the heterogeneity observed in the epidemiological outcomes of the pandemic. This reinforces the need for approaches that systematically consider socioeconomic determinants.
In this context, using generalized models, we identified associations between socioeconomic factors and vaccination status for both COVID-19 and influenza, as well as the influence of socioeconomic conditions on mask-wearing and social distancing. Importantly, we also observed differences in contact behaviors based on employment status while education level did not show a significant association. 
These findings highlight the complex interplay of socioeconomic and demographic factors in shaping individual responses to public health measures. Understanding these dynamics is essential for developing effective epidemic models and targeted public health strategies, particularly for vulnerable populations. 
\end{abstract}

\section*{Introduction}
Socioeconomic status (SES) is a critical determinant of health outcomes, with far-reaching implications for both communicable and non-communicable diseases \cite{mamelund_association_2021,foster_understanding_2021}. 
While a substantial research effort has been devoted to examining the relationship between socioeconomic disparities and the prevalence of non-communicable diseases  \cite{allen_socioeconomic_2017,stringhini_socioeconomic_2017,sommer_socioeconomic_2015,mtintsilana_association_2023, foster_understanding_2021,lantz_socioeconomic_1998}, it is also evident that socioeconomic divides play a pivotal role in the transmission of infectious diseases. The COVID-19 pandemic further emphasized the role of socioeconomic factors in determining health outcomes, reinforcing their significance even in the context of communicable diseases. 
Protective behaviors, both pharmaceutical and non-pharmaceutical, were seen to be mediated by socioeconomic factors.
Indeed, while, non-pharmaceutical interventions (NPIs) such as social distancing were instrumental in reducing infection rates and alleviating the burden on the healthcare system \cite{wellenius_impacts_2021,duque_timing_2020}, their adoption varied significantly across socioeconomic groups. Moreover, vaccination uptake was found to correlate with education level and household income in some countries \cite{reichmuth_socio-demographic_2023}. These behavioral disparities may have contributed to varying epidemic outcomes across socioeconomic groups. In fact, in several countries, populations in lower SES areas experienced faster virus spread and transmission, putting them at greater risk of COVID-19, infection, hospitalization, and mortality compared to their higher SES counterparts \cite{fernandez-martinez_socioeconomic_2022,zhang_fine-scale_2022}. \\ 
In line with previous studies, Italy's experience with the COVID-19 pandemic has shown a varied impact across different socioeconomic strata right from the early stages  \cite{consolazio2021assessing, buja_demographic_2020}. During the lockdown and post-lockdown periods, COVID-19 case rates were higher in more deprived municipalities compared to less deprived ones. However, hospitalization and case fatality rates did not show significant differences based on deprivation levels during any of the studied periods \cite{mateo-urdiales_socioeconomic_2021}.
In response to the pandemic, Italy implemented stringent measures and a regional tiered system of non-pharmaceutical interventions (NPIs)  \cite{DPCM031120}, which influenced population contact behaviors \cite{Manica2021} and, consequently, the rate of COVID-19 transmission \cite{tizzani_impact_2023}. 
Even with uniform measures across the country, there was a heterogeneous behavioral response across socioeconomic groups indicating that socioeconomic characteristics shaped individual and community-level responses to these interventions \cite{gauvin2021socio}.

These observations during the pandemic stress the importance of analyzing behavioral disparities across socioeconomic groups to eventually ensure equitable pandemic preparedness and response.\\
To gain a deeper understanding of the specific SES factors that influence epidemic dynamics, further research is necessary, particularly to incorporate external factors, such as socioeconomic conditions, into the primary driver of communicable diseases: contact behavior. The epidemiological significance of age-stratified contact patterns in evaluating the impact of human behavior on epidemics has been extensively investigated prior to the COVID-19 pandemic \cite{mossong2008social, Fumanelli2012}. Additionally, this methodology has been employed in multiple studies on the effect of contact behavior during the COVID-19 pandemic \cite{coletti2020comix, wong_social_2023, liu_rapid_2021}, suggesting the need for updated contact surveys during a global crises. Age has been demonstrated to encapsulate a significant proportion of the observed inter-individual variation in contact patterns, and empirical age-dependent contact matrices have been measured in various settings and used for epidemic modeling. Additionally, contact matrices, which can account for the stratification of contacts by factors such as age, location, socioeconomic conditions, and time, can offer a deeper quantitative understanding of epidemic dynamics \cite{manna_generalized_2023}. 
However, while socioeconomic factors significantly influence epidemic outcomes, research on their effects on contact behavior remains limited. In this direction, the implications of socioeconomic status (SES) on epidemic modeling have been explored during the COVID-19 pandemic in Hungary \cite{manna_importance_2024}. Additionally, studies in the UK, Netherlands, Switzerland, and Belgium have assessed the change in mean contact numbers stratified by demographic factors, including SES, in the post-pandemic period \cite{jarvis_social_2024}. Such a results calls for additional evaluation of contact patterns after the pandemic. Moreover, a gap remains in understanding the associations between socioeconomic status (SES) and the adoption of protective behaviors after the state of emergency was lifted during the COVID-19 pandemic.\\
To answer these research gaps, we conducted a contact survey representative of the Italian population, stratified by demographic and socioeconomic indicators four years after the start of the COVID-19 pandemic.
First, to have a broader picture of the heterogeneous impact of the COVID-19 pandemic on the well-being of participants across different socioeconomic strata, we examine their perceptions of the impact on their social, psychological, economic, and work status.
Second, we assessed the associations of SES with contact and protective behaviors adopted by participants at the time of the study. This included their adherence to non-pharmaceutical preventive measures, and the contact behavior of participants according to the various demographic and socioeconomic stratification of the population. This approach enabled us to gain a deeper understanding of the interrelationship between individual characteristics, protective behaviors, and social interactions in the context of the post-pandemic period. 
By examining the links between SES, protective behaviors, and contact patterns we aim to provide valuable insight to better inform epidemic models and develop equitable strategies for future pandemic preparedness. In particular, we aim to identify significant socioeconomic determinants of protective and contact behaviors, key factors for the spread of communicable diseases.

\section*{Data and Methods}

\subsection*{Data description}
To assess and analyze disparities in protective and contact behaviors in the aftermath of the COVID-19 pandemic, we designed a survey targeting the Italian population.
The survey, representative of the Italian population stratified by \textit{sex}, \textit{age}, \textit{geographical area}, \textit{education}, \textit{city size}, and \textit{employment status}, was administered in March 2024 through computer-aided web interview technique (CAWI). A sample of $N=1200$ participants was randomly extracted from the proprietary web panel from Doxa S.p.a together with their stratification weights. Weights are based on the Italian National Institute of Statistics ISTAT data as of December 31, 2022, for population demographics and December 31, 2021, for occupational status.   Questions asked of the participants and descriptive statistics are reported in the supplementary material \nameref{SI.0}, and \nameref{tabsi:descr_stat} respectively.\\
On the demographic and socioeconomic side, the survey collected information on participants' sex (male or female), age, geographical location (\textit{northwest}, \textit{northeast}, \textit{center}, \textit{southern}, and \textit{islands} Italian regions), city size ($0-10^3$, $10^3-10^6$, and more than $10^6$ inhabitants), employment status (\textit{employed} or \textit{unemployed}), and education level (university degree or not). \\
First, to evaluate the  heterogeneity of the perceived impact of the pandemic, we asked participants whether they felt it had negatively affected their lives economically, professionally, socially, and mentally. \\
Turning to protective health behaviors, the survey assessed participants' vaccination status for COVID-19 and influenza, asking whether they were vaccinated. This allowed us to gauge vaccination uptake across different population segments. Furthermore, we explored participants' attitudes and adherence to NPIs measures, such as mask-wearing and social distancing. 

Lastly, the survey's main objective was to collect participants' contact behavior patterns. Participants
were asked on a Tuesday to record the number of contacts they had during the latest working and
non-working days, namely Sunday and Monday. They categorized these contacts as either individual (i.e., people they met in person with whom they exchanged at least a few words or had physical contact) or collective (i.e., individuals within a 1.5-meter radius with whom they did not exchange words or physical contact). For individual contacts, participants were also asked to provide the ages of those they interacted with. Additionally, they were asked to record the contact location from the following options \textit{home}, \textit{work}, \textit{essential activities} (i.e. grocery shopping), \textit{leisure activities} (i.e. sports), \textit{transport}, \textit{health} (i.e. medical examination), \textit{study}. \\

Additionally, to qualitatively compare contact patterns before, during, and after the pandemic, we gathered data from previous contact surveys. We used the POLYMOD survey conducted in 2008 to represent the pre-pandemic period \cite{mossong2008social}, and the COMIX survey \cite{coletti2020comix} from the second wave of COVID-19 in Italy to represent the pandemic period. Both of these prior surveys collected comparable demographic information and age-stratified contact data for representative samples of the Italian population at different times.

\subsection*{Statistical model of participants' perception and behavior}
To analyze the association between individual-level factors and participants' behavior, we leveraged generalized linear models.
The explanatory variables encompassed both demographic and socioeconomic characteristics. On the demographic front, we considered factors such as \textit{age}, \textit{gender}, \textit{city size}, and \textit{geographic area}. These variables allowed us to capture the potential heterogeneity in experiences and behaviors across different population segments.
We also incorporated socioeconomic determinants, including \textit{employment status} and \textit{education level}. All the explanatory variables were categorical : age was stratified into five groups (18-24, 25-34, 35-44, 45-54, 55+), the gender was divided into male and female, the city size was layered by the number of inhabitants ($0-10^3$, $10^3-10^6$, and more than $10^6$ inhabitants), the geographic area (\textit{geo\_area}) was stratified into four, the \textit{northwest}, \textit{northeast}, \textit{center}, \textit{southern} and \textit{islands} Italian regions, employment was divided between \textit{employed} and \textit{unemployed} participants, and education was divided between participants with a university degree (\textit{degree}) and participant without (\textit{nodegree}).\\

We focused on three types of outcome variables: participants’ perceptions of the pandemic’s impact, the protective behaviors they adopted, and the number of contacts they had, distinguishing between individual and collective contacts. We implemented the Brant test \cite{brant_assessing_1990} to assess the proportional odds assumption for the ordinal model. The results of the Brant test indicated no evidence against the proportional odds assumption, suggesting that the ordinal regression model was reasonable for our data. Details on how each type of outcome variable was incorporated into the models are described below. 

\subsubsection*{Perception of the Impact of COVID-19}
To assess the perception of the impact of COVID-19 on the participants we employ a multivariate ordinal logistic regression. The participants were asked whether they agreed with the statement that COVID-19 had a negative impact on their well-being in the economic, work, psychological, and social conditions. In this case, the ordinal outcome $Y$ of the model is the participants' answer measured on the Likert scale,  $Y\in \{\textit{disagree} < \textit{neutral} < \textit{agree}\}$. Formally, considering $P(Y\leq \textit{k})$ the cumulative probability of the event $Y\leq k$, with $k = 1, 2$, represents the level disagreement with the statement, the ordinal linear regression model can be written as

\begin{equation}logit(P(Y\leq \textit{k}))=log \left(\frac{P(Y\leq \textit{k})}{P(Y> \textit{k})} \right) = \beta_{k, 0}+\mathbf{X}^T \boldsymbol{\beta}\end{equation}

where \begin{equation}\boldsymbol{\beta}=\begin{bmatrix}
  \beta{\text{age}} \\ \beta{\text{gender}}\\ \beta{\text{city\_size}}\\ \beta{\text{geo\_area}}\\ \beta{\text{employment}}\\ \beta{\text{education}}
\end{bmatrix}
\label{beta_vector}
\end{equation}
is the vector of coefficients, and 

\begin{equation}\mathbf{X} = \begin{bmatrix}
   x_{\text{age}} \\ x_{\text{gender}}\\ x_{\text{city\_size}}\\ x_{\text{geo\_area}}\\ x_{\text{employment}}\\ x_{\text{education}}
\end{bmatrix}
\label{x_vector}
\end{equation}
is the vector of explanatory variables. In what follows, we will use the same explanatory variables as in Eq.\ref{x_vector} unless specified otherwise. 
The analysis was performed using the MASS library for R \cite{noauthor_modern_nodate}.
\subsubsection*{Protective behavior}
Participants' behavior toward protective measures was evaluated through questions about the number of vaccination intakes for COVID-19 and influenza, mask-wearing attitude, and actuation of social distancing. 
Social distancing and mask-wearing attitudes were assessed with an ordinal logistic regression model. The ordinal outcome $Y$ was measured on a Likert scale, $Y\in \{\textit{never} < \textit{sometimes} < \textit{always}\}$. 
Formally, considering $P(Y\leq \textit{k})$ the cumulative probability of the event $Y\leq k$, where $k = 1, 2$, the ordinal linear regression model can be written as

\begin{equation}logit(P(Y\leq \textit{k}))=log \left(\frac{P(Y\leq \textit{k})}{P(Y> \textit{k})} \right) = \beta_{k, 0}+\mathbf{X}^T \boldsymbol{\beta}\end{equation}

Vaccination uptake was modeled with a logistic regression where the number of vaccination doses was mapped into a binary variable $Y$ (\textit{yes} or \textit{no}) and the model is expressed by equation
\begin{equation}logit(P(Y=\textit{yes}))=log(\frac{P(Y= \textit{yes})}{P(Y =\textit{no})})= \beta_{yes, 0}+\mathbf{X}^T \boldsymbol{\beta}\end{equation}
were $\boldsymbol{\beta}$ and $\mathbf{X}$ are expressed by Equations \ref{beta_vector} and \ref{x_vector}.

\subsubsection*{Contact behavior}
To understand participant contact behavior we consider a negative-binomial Bayesian model \cite{manna_social_2023}, the dependent variable in this case was the total number of contacts $Y_
{i}$ for each participant $i$ and the model can be formally expressed by
\begin{equation}Y_i\sim \textit{NegBin}(\mu_i,r)\end{equation}
and the link function
\begin{equation}\log(\mu_i)= \beta_{0}+\mathbf{X}^T \boldsymbol{\beta}\end{equation}

were $\boldsymbol{\beta}$ and $\mathbf{X}$ are defined by Equations \ref{beta_vector} and \ref{x_vector}, $\mu_i$ the mean number of contacts for each participant $i$, and $r$ is the reciprocal dispersion parameter: the highest $r$ is the closest $Y_i$ is Poisson distributed. We considered weakly informed priors (Gaussian distributed) for the coefficients $\boldsymbol{\beta}$.  
To gain a better understanding of the factors influencing contact behavior, we conducted a stratified analysis by dividing the data into weekday (Monday) and weekend (Sunday). This allowed us to investigate potential differences in contact patterns based on working and non-working participants. To discern the working behavior from the employment status we integrated information on whether the participant had worked on the respective day into the employment variable. This resulted in considering three categories instead of one in the explanatory variable about the employment status: \textit{working-employed}, i.e. participants who were employed and worked on the day of interest (weekday or weekend), \textit{non-working-employed}, i.e. employed participants who did not work on the day of interest, and \textit{unemployed} participants. Notice the intercept of the model is given by the following categories: age[18-24], education[degree], city size[0-10000], country area[Center], and employment[unemployed].\\
By incorporating this additional layer into the employment status, we were able to better capture the potential influence of work-related activities on contact behavior.
Furthermore, we built separate models for each specific contact location, such as essential and leisure activities, transport, or health. By selecting only the contacts occurring in a particular setting, we could explore the differences in the determinants of contact behavior within that specific location and understand whether heterogeneity in contact patterns varies across the different settings.
The analysis was performed with Bambi (BAyesian Model Building Interface) \cite{capretto_bambi_2022}, a Python library for Bayesian modeling.

\section*{Results}

\subsection*{Impact of COVID-19 on the perception of well-being}

\begin{figure}[!htb]
  \centering
  \includegraphics[width=1\linewidth]{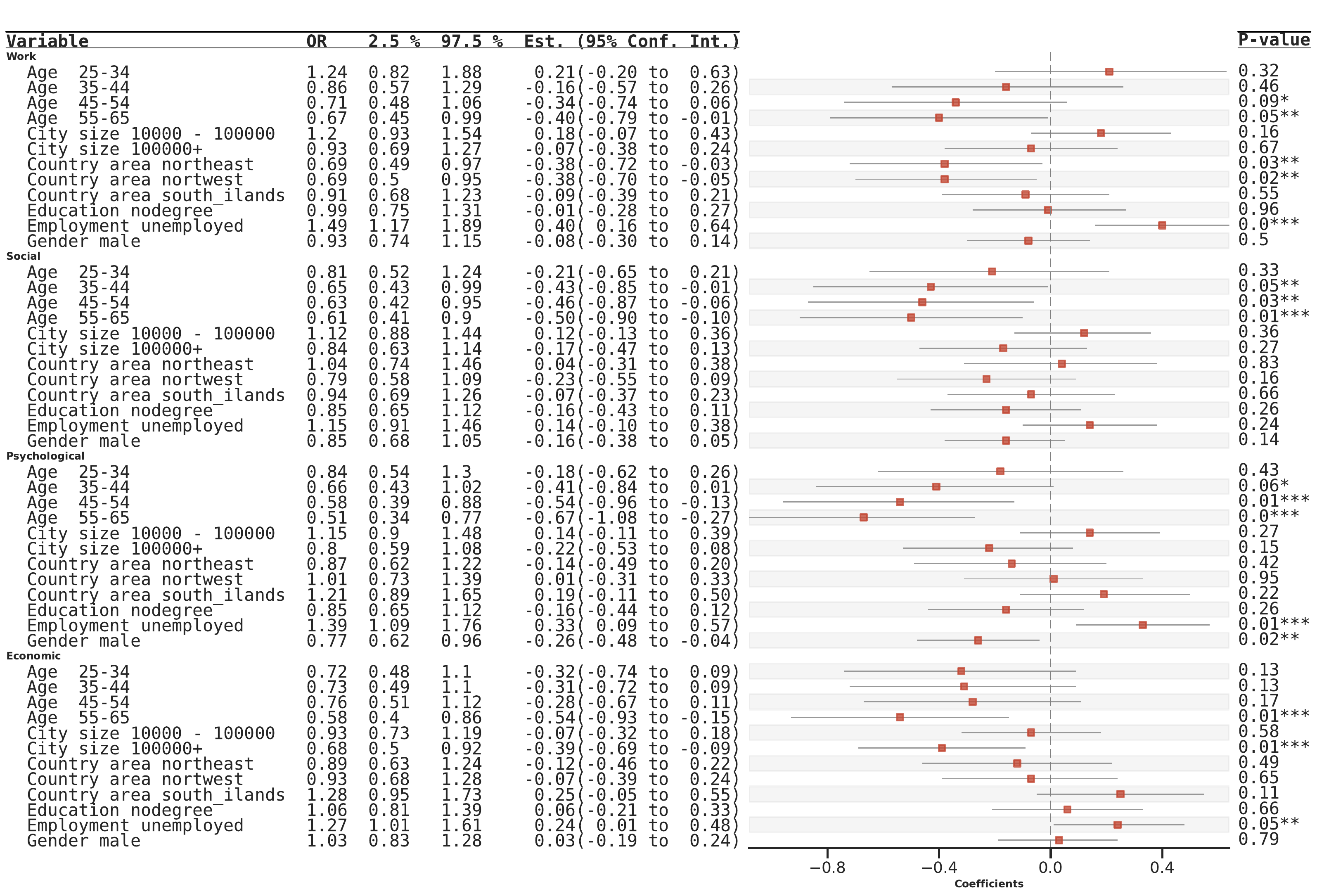}
  \caption{Ordinal logistic regression model for the perception of the impact of COVID-19 on work, social life, psychological well-being, and economic situation. P-value annotation legend: *: $1.00\times 10^{-2} < p < 5.00\times 10^{-2}$ , **: $1.00\times 10^{-3} < p <= 5.00\times 10^{-2}$ , ***: $1.00 \times 10^{-4} < p <= 1.00\times 10^{-3}$.
  The table shows the names of the explanatory variables (Variable), the odd ratio (OR), the confidence interval for the odd ratio (2.5\% and 97.5\%), and the estimated values of the model with their confidence interval (Est.)}
  \label{fig:olm_perception}
\end{figure}

In this section, we examine participants' perceptions of the impact of COVID-19 across four key aspects of their lives: professional life, social interactions, psychological and economic well-being. This analysis aims to assess the heterogeneity of COVID-19's effects in the post-pandemic context, extending beyond the variations observed in the pandemic's epidemiological outcomes.
Table \ref{tab:part_pc} shows the total number and the percentage of the participants' answers about their perception of the negative impact of the pandemic from an economic, psychological, social or professional point of view. Overall, approximately 20\% of participants reported being neutral over the perception of the negative impact of the pandemic. 
Around 40\% of the participants recognized a negative influence on their economic, social, and psychological well-being, while around 30\% of the participants disagreed with the statement.
To get an overview of the underlying socioeconomic and demographic factors behind the perception of the impact of COVID-19 on participant well-being, in Figure \ref{fig:olm_perception} we show the results of the four ordinal logistic regression. Notice that the percentages in the description of the results are computed from the odd ratio derived from the statistical model. The reference group for the analysis comprised participants with the following characteristics: aged 18-24 years old, female, residing in cities with populations under 10,000 inhabitants, living in central regions, holding a university degree, and employed. This reference group served as the baseline against which the effects of different socioeconomic and demographic factors were compared in the analysis,  isolating the independent influence of each variable on outcomes related to the perception of pandemic impact, and protective behaviors.

\begin{table}[H]
\let\center\empty
\let\endcenter\relax
\centering
\resizebox{.9\width}{!}{\begin{tabular}{|l|l|r|r|}
\hline
 &  & Count & Percentage (\%) \\
Question & Answer &  &  \\
\hline
\multirow[t]{3}{*}{Economic Impact} & agree & 479 & 39.92 \\
 & disagree & 401 & 33.42 \\
 & neutral & 320 & 26.67 \\
\cline{1-4}
\multirow[t]{3}{*}{Psychological Impact} & agree & 550 & 45.83 \\
 & disagree & 395 & 32.92 \\
 & neutral & 255 & 21.25 \\
\cline{1-4}
\multirow[t]{3}{*}{Social Impact} & agree & 538 & 44.83 \\
 & disagree & 383 & 31.92 \\
 & neutral & 279 & 23.25 \\
\cline{1-4}
\multirow[t]{3}{*}{Work Impact} & agree & 359 & 29.92 \\
 & disagree & 591 & 49.25 \\
 & neutral & 250 & 20.83 \\
\cline{1-4}
\hline
\end{tabular}
}
\caption{Descriptive statistics of the answers to the participant's perception of the negative impact of COVID-19.}
\label{tab:part_pc}
\end{table}
\subsubsection*{Impact on work}
The perception of the negative impact of the COVID-19 pandemic on work was found to vary significantly by age, geographical location, and employment status. In particular, older participants in the age group 55-65 were 32,8\% more likely to not agree compared to the younger participants (age group 18-24), (OR=0.67, CI=0.45-0.99). Similarly, participants living in northern regions were 31.4\% (OR=0.69, CI=0.49-0.97) more likely to disagree than participants living in central regions. Conversely, unemployed participants were more likely to agree than employed participants (OR=1.47, CI=1.17-1.89).
These findings suggest that age, geographical location, and employment status were significant factors influencing how individuals perceived the pandemic's impact on their work situation. Older adults and those living in northern regions, were more likely to disagree with the perception of the negative impact of the pandemic, while unemployed individuals appeared to have relatively more negative perceptions of the work-related effects of the COVID-19 crisis.
\subsubsection*{Impact on social life}
Perceptions of the impact of the pandemic on social life varied only significantly by age group. Compared to younger participants (18-24 years), participants older than 34 years old were more likely to disagree with the negative effect on their social life. In particular, those aged 35-44 were 34.8\% more likely to disagree (OR=0.65, CI=0.43-0.99), those aged 45-54 were 36.9\% more likely to disagree (OR=0.63, CI=0.42-0.95) and those aged 55-64 were 39.3\% more likely to disagree (OR=0.61, CI=0.41-0.9). The other explanatory variables were not significant.
Overall, older age groups were significantly more likely to disagree with perceptions of the pandemic's impact on social life compared to youngest participants (18-24 years), while other factors like geography and employment status did not have a significant effect.
\subsubsection*{Impact on psychological well-being }
Perceptions of the  negative impact of the pandemic on psychological well-being varied by age group, gender, and employment status. Compared with younger participants (18-24 years old), participants older than 44 years old were more likely to disagree with the negative effect on their psychological well-being. In particular, participants in the 45-54 age group and those aged 55-65 were 41.8\% (OR=0.58, CI=0.39-0.88) and 48.8\% (OR=0.51, CI=0.34-0.77) more likely to disagree respectively. Male participants were 22.9\% more likely to disagree than female participants (OR=0.77, CI=0.62-0.96) and unemployed participants were more likely to agree than employed participants (OR=1.39, CI=1.09-1.76).
Perceptions of the pandemic's impact on psychological well-being were influenced by age, gender, and employment status, with younger, female, and unemployed participants being more affected.
\subsubsection*{Impact on economic}
Perceptions of the impact of the pandemic on economic status varied by age group, city size, and employment status. Compared with younger participants (18-24 years old), those aged 55-65 were 41.6\% more likely to disagree with the statement (OR=0.58, CI=0.4-0.86), and those living in densely populated cities were 32\% more likely to disagree than those living in less densely populated areas (OR=0.68, CI=0.5-0.92). Conversely, unemployed participants were more likely to agree (OR=1.27, CI=1.01-1.61).
Perceptions of the pandemic's economic impact differed by age, city population density, and employment status, with older participants and participants living in more densely populated areas feeling less affected by the pandemic. However, unemployed individuals perceived a stronger economic impact of the pandemic than employed participants.

\subsubsection*{Socioeconomic and demographic effects on the perception of the COVID-19 Impact}
Demographic and socioeconomic factors largely influenced the perception of the negative impact of the COVID-19 pandemic. 
Age emerged as a significant factor, with younger participants reporting experiencing the pandemic more intensely. The perceived impact appeared to diminish with increasing age, suggesting that older individuals may have been better equipped to cope with the challenges posed by the crisis \cite{Ceccato2020}.
Employment status also played a significant role, as unemployed participants reported a more pronounced negative impact of the pandemic compared to their employed counterparts on the professional, economic, and psychological points of view. 
Geographical differences were also observed, with participants in the northern regions perceiving a less pronounced impact on their work compared to those in the central regions. These results may reflect the intrinsic regional differences in the economic and social stratification that might have been exacerbated by the pandemic.
Gender also emerged as a factor influencing the perception of the psychological burden, with female participants reporting a higher level of agreement on the pandemic's detrimental impact on their well-being. This suggests that women may have experienced greater mental health challenges during the crisis, confirming previous results during the pandemic times \cite{higham_epidemics_2021, dotsikas_gender_2023}. 
Furthermore, the perception of the economic burden was influenced by city size, with those residing in larger urban centers reporting a lesser impact of the pandemic. This may be attributed to the greater economic resilience and resources available in larger metropolitan areas \cite{arin_urban_2022}.
Collectively, these findings highlight the significant socioeconomic disparities in how individuals perceive the impacts of the COVID-19 pandemic, adding to the already observed disparities in the epidemic outcomes. This shows the importance of understanding socioeconomic differences even before a potential pandemic. Specifically, it is essential to investigate variations in protective and contact behaviors across socioeconomic statuses (SES). Such an investigation can shed light on the extent to which the observed disparities in pandemic outcomes are driven by differences in these behaviors across socioeconomic status (SES) groups..

\subsection*{Protective behaviour}
\begin{figure}
  \centering
  \includegraphics[width=1\linewidth]{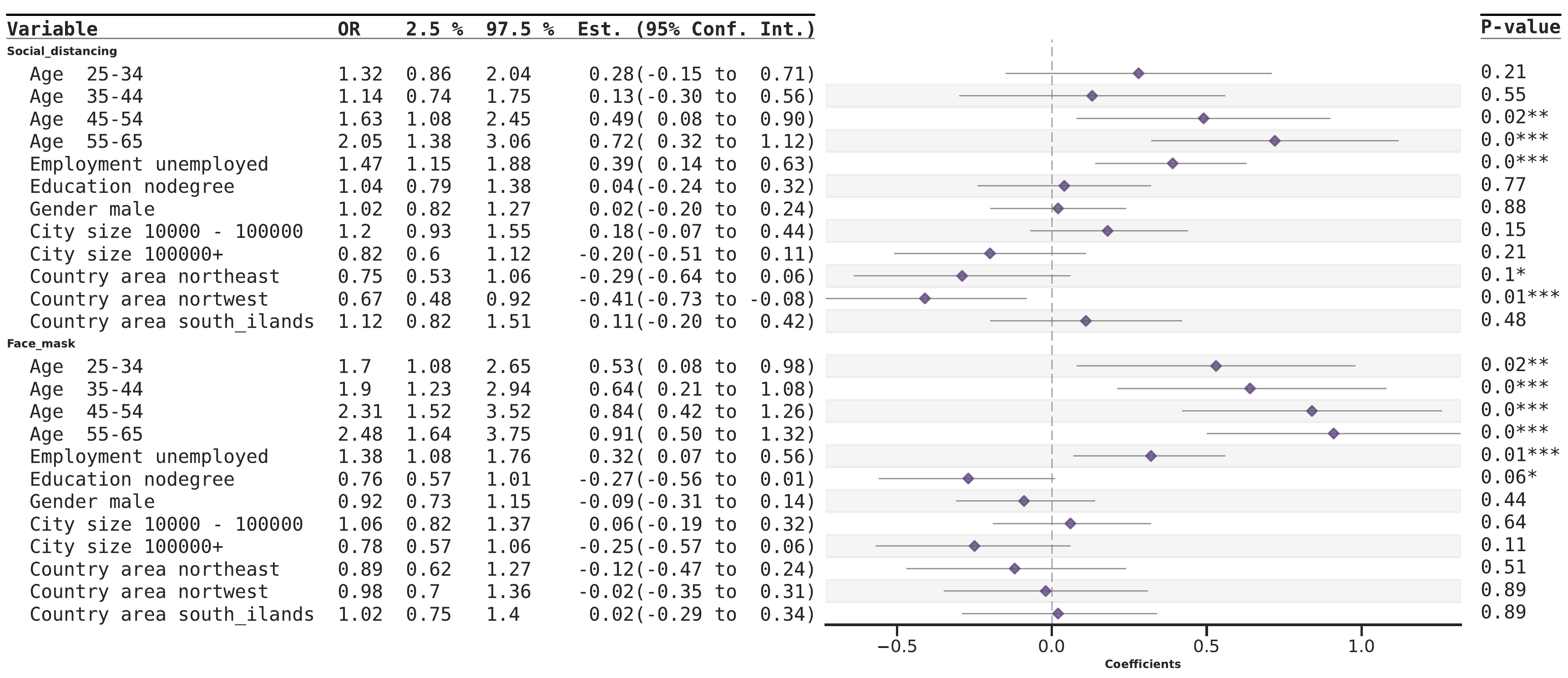}
  \caption{Ordinal logistic regression model for the non-pharmaceutical protective behavior. P-value annotation legend: *: $1.00\times 10^{-2} < p < 5.00\times 10^{-2}$ , **: $1.00\times 10^{-3} < p <= 5.00\times 10^{-2}$ , ***: $1.00\times 10^{-4} < p <= 1.00\times 10^{-3}$. The table shows the names of the explanatory variables (Variable), the odd ratio (OR), the confidence interval for the odd ratio (2.5\% and 97.5\%), and the estimated values of the model with their confidence interval (Est.)}
  \label{fig:prot_beh}
\end{figure}

\begin{figure}[!htb]
  \centering
  \includegraphics[width=1\linewidth]{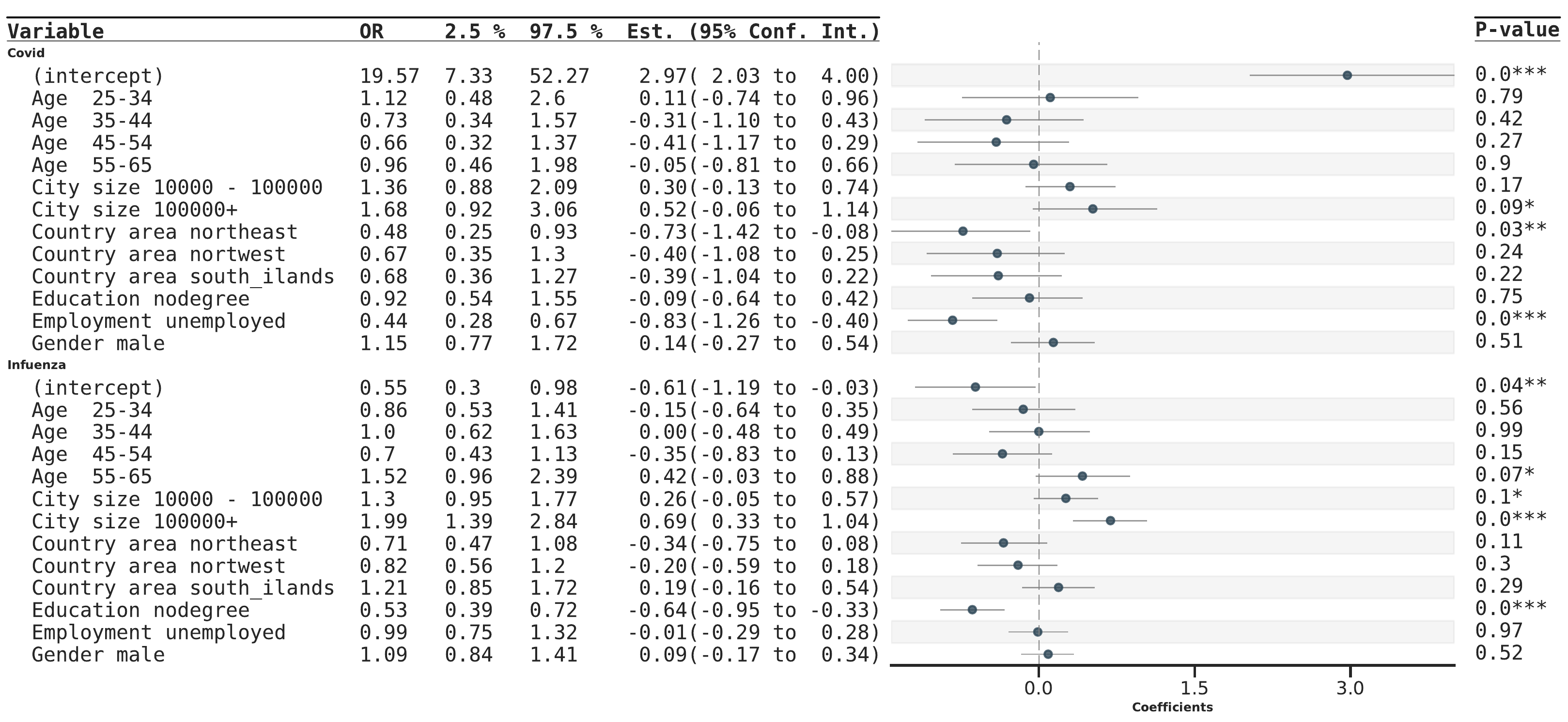}
  \caption{Binomial regression model for vaccination uptake. P-value annotation legend: *: $1.00\times 10^{-2} < p < 5.00\times 10^{-2}$ , **: $1.00\times 10^{-3} < p <= 5.00\times 10^{-2}$ , ***: $1.00\times 10^{-4} < p <= 1.00\times 10^{-3}$.
  The table shows the names of the explanatory variables (Variable), the odd ratio (OR), the confidence interval for the odd ratio (2.5\% and 97.5\%), and the estimated values of the model with their confidence interval (Est.)}
  \label{fig:prot_vaccine}
\end{figure}

We assessed protective behaviors by asking participants about their adherence to non-pharmaceutical interventions, such as social distancing and face mask-wearing, as well as their vaccination uptake for both COVID-19 and influenza. 
\\
Table \ref{tab:part_pb} presents the total number and percentage of participants who adopted various protective behaviors. Compatible with the official report from the Italian government \cite{governoMinisteroDella} participants (over 90\%) reported being vaccinated against COVID-19, likely due to the mandatory policies for healthcare workers and strongly encouraged policies for the rest of the population in place during the pandemic. Around 30\% of the participants had received the influenza vaccine, 10\% higher than the vaccine uptake for the general population reported by the Italian minister of health during the 2022-2023 flu season \cite{saluteDatiCoperture}.
The adoption of non-pharmaceutical interventions (NPIs) like mask-wearing and social distancing was less consistent. The majority of participants either sporadically or never wore face masks, with about 10\% reporting always using masks. Similarly, around 15\% consistently practiced social distancing, while most did so occasionally or not at all.
To further investigate the heterogeneity across demographic and socioeconomic strata we implement an ordinal logistic regression model for non-pharmaceutical intervention, shown in Figure \ref{fig:prot_beh}, and a binomial regression model for vaccine uptake, Figure \ref{fig:prot_vaccine}. Notice that the percentages in the description of the results are computed from the odd ratio derived from the statistical model.
\begin{table}[H]
\let\center\empty
\let\endcenter\relax
\centering
\resizebox{.9\width}{!}{\begin{tabular}{|l|l|r|r|}
\hline
 &  & Count & Percentage (\%) \\
Question & Answer &  &  \\
\hline
\multirow[t]{2}{*}{Covid vaccine} & No & 113 & 9.42 \\
 & Yes & 1074 & 89.50 \\ 
\cline{1-4}
\multirow[t]{2}{*}{Flu vaccine} & No & 818 & 68.17 \\
 & Yes & 366 & 30.50 \\
\cline{1-4}
\multirow[t]{3}{*}{Face mask} & Always & 119 & 9.92 \\
 & Never & 510 & 42.50 \\
 & Sometimes & 571 & 47.58 \\
\cline{1-4}
\multirow[t]{3}{*}{Social dist} & Always & 191 & 15.92 \\
 & Never & 534 & 44.50 \\
 & Sometimes & 475 & 39.58 \\
\cline{1-4}
\hline
\end{tabular}
}
\caption{Descriptive statistics of the answers to the participants' protective behavior.}
\label{tab:part_pb}
\end{table}
\subsubsection*{Social distancing}
Older participants were more likely to always adhere to social distancing guidelines. In particular, participants within the age group 45-54 were more likely to comply with the guidelines \cite{saluteComeProteggersi}, compared to younger participants (age group 18-24) (OR=1.63, CI=1.08-2.45). Furthermore, participants in the age group 55-65 were more likely to always adopt social distancing (OR=2.05, CI=1.38-3.06). Unemployed individuals were also more likely to adopt social distancing guidelines consistently than employed participants (OR=1.47, CI=1.15-1.88). Conversely, those living in northwestern regions were 33.4\% less likely to comply with social distancing guidelines than those residing in central regions (OR=0.67, CI=0.48-0.92).
\subsubsection*{Face mask}
Across all age groups, participants were more likely to consistently wear face masks compared to younger individuals (aged 18-24). This effect was particularly pronounced in older participants. Additionally, unemployed individuals were more likely to consistently wear face masks compared to employed individuals (OR=1.38, CI=1.08-1.76).

\subsubsection*{Vaccine uptake}
COVID-19 vaccine uptake was lower for individuals living in the northeast of Italy (OR=0.48, CI=0.25-0.93) compared to those living in central regions, as well as for unemployed participants (OR=0.44, CI=0.28-0.67) compared to unemployed. Conversely, influenza vaccine uptake was higher for those living in highly populated areas (OR=1.99, CI=1.39-2.84), while participants without a degree were less likely to be vaccinated (OR=0.53, CI=0.39-0.72). 
It is important to note that the COVID-19 vaccine was mandatory for healthcare workers until November 2022, while for the rest of the Italian population, non-vaccination was discouraged by limited access to public enclosed spaces until January 2023. We did not control for the occupation type of the participant, so the effect of being a healthcare worker might not be captured by the analysis. 
\subsubsection*{Socioeconomic and demographic factors of protective behavior}
Overall participants' ages were indicative of adherence to non-pharmaceutical interventions. Older participants were more likely to consistently adopt face masks and social distancing measures compared to younger participants. On the other hand, the analysis suggests that age did not play a role in the vaccine uptake. 
Employment status was also found to be associated to adopting NPIs and vaccination. Unemployed participants were more likely to adopt NPIs consistently, such as face masks and social distancing, but less likely to be vaccinated against COVID-19. 
The geographic region also influenced vaccination rates. Participants from northeastern regions were less likely to consistently adopt social distancing and less likely to be vaccinated against COVID-19 compared to central regions. 
Education level was another significant factor influencing vaccination decisions. Participants without a degree were less likely to be vaccinated against influenza.
The results demonstrate that demographic, socioeconomic, and geographic characteristics play a significant role in shaping how individuals adopt protective behaviors four years after the pandemic. This highlights the need to account for these factors when developing epidemiological models, in order to more accurately capture disease transmission patterns and dynamics.

\subsection*{Contact behavior and socioeconomic variables}
Modeling the contact patterns of the population is crucial for understanding the transmission dynamics of communicable diseases like COVID-19. To this end, we conducted a two-fold analysis of the participants' contact behavior. 
First, we provided an overview of the age-stratified contact matrices, which quantify the average number of contacts between individuals of different age groups. This descriptive analysis offered an overview of the overall structure and patterns of social interactions within the population after the pandemic.
Second, we employed the negative binomial model to investigate the factors influencing contact behavior. By incorporating socioeconomic and demographic variables into our models, we identified the key determinants of individual contact patterns. With this approach, we explored how factors such as age, employment status, education level, and geographic location shape social interactions.
\subsubsection*{Contact patterns before, during, and after the COVID-19 pandemic}
\begin{figure}[!h]
  \centering
  \includegraphics[width=1\linewidth]{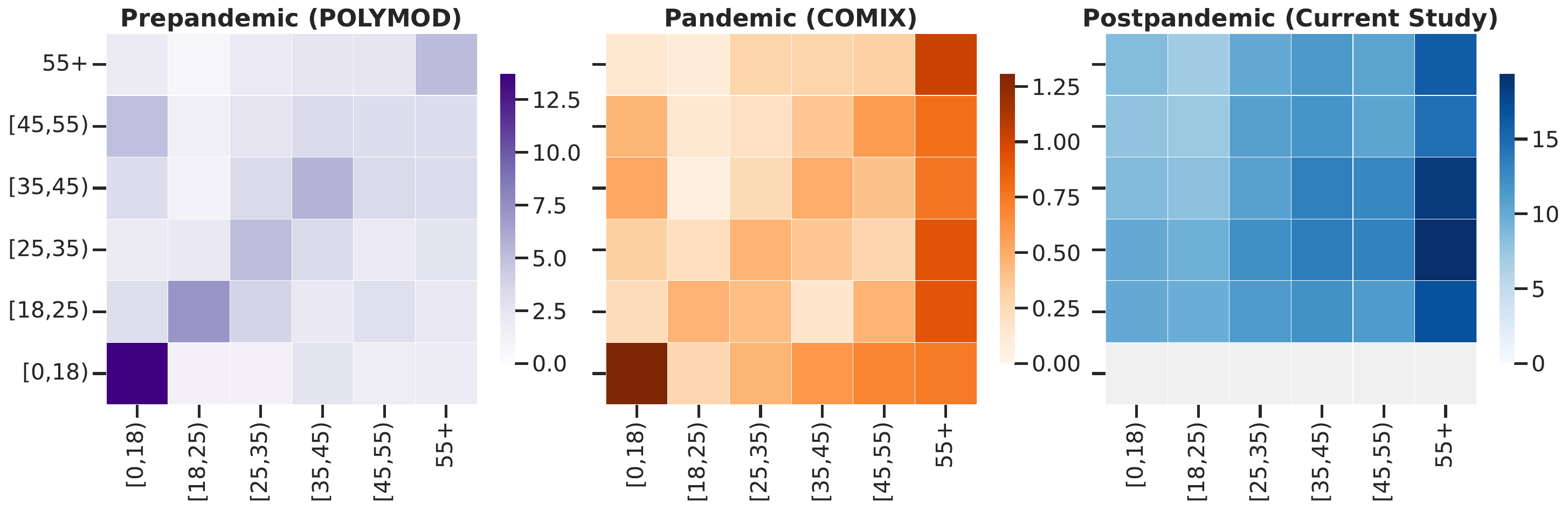}
  \caption{Contact patterns before COVID-19 (Prepandemic), during COVID-19 (Pandemic), and after COVID-19 (Post-pandemic). The matrix element ($m_{ij}$) is the median of the bootstrap realizations of the average number of contacts of age group $i$ with participants of age group $j$. For each matrix, we consider only individual contacts over the whole period of the study.}
  \label{fig:cm_comparison}
\end{figure}

To assess the differences in contact patterns across the pre-pandemic, pandemic, and post-pandemic periods we compare our study with the two contact surveys POLYMOD and COMIX. In Figure \ref{fig:cm_comparison} we show the contact patterns of the three periods. The matrix element is defined as the median of 10,000 bootstrap realizations of the mean number of individual contacts for each age class of participants in contact with other age classes. Further details on the methods employed can be found in the supplementary information  \nameref{SI.2},  with particular reference to Figure \siref{fig:cm_comparison}, which reports the explicit values of the median and interquartile range for each element.
The homophily hypothesis of contact patterns can be spotted in all three periods, with the difference that during and after the pandemic off-diagonal elements are more populated than the pre-pandemic ones. In particular, younger participants have more contact across all age groups, this effect is stronger in the post-pandemic period. Let us notice that differently from POLYMOD and COMIX our survey does not cover participants older than 65 years old and younger than 18 years old, and the survey methodology was different. 

\subsubsection*{Contact patterns}
\begin{figure}[!hb]
  \centering
  \includegraphics[width=1\linewidth]{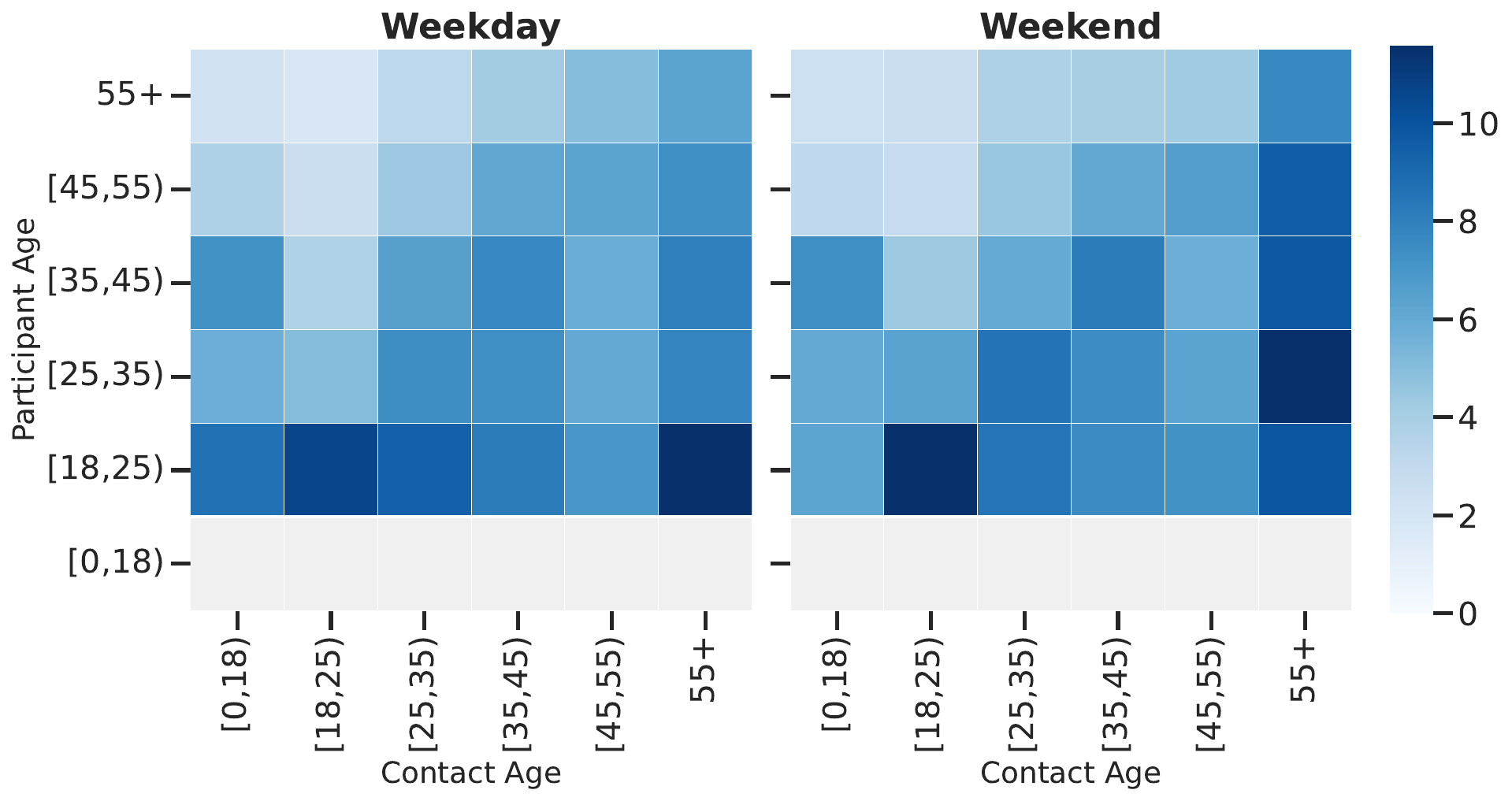}
  \caption{Contact matrices on weekday and weekend. The matrix element ($m_{ij}$) is the median of the bootstrap realizations of the average number of contacts of age group $i$ with participants of age group $j$.}
  \label{fig:cm_general}
\end{figure}

In the following analysis, we consider the individual contact patterns of the participants on weekdays and weekends, first without differentiating with respect to the contact location, and subsequently, considering different contact locations. In Figure \ref{fig:cm_general} we show the contact matrices for the two days of the surveys. Overall contact patterns show similar trends on both days, with more contacts on average during the weekend for younger participants, and more contacts within the age group 0-18 for participants between 35 and 45 years old. Younger participants have more interactions overall, in particular among the same age group, but also with the oldest age group. 

Considering different contact locations, Fig. \nameref{SI2.1} and \nameref{SI2.2}, overall younger participants had more contacts than older ones in all the locations both on weekdays and weekends. In particular, contacts on the weekday during essential activities are heterogeneous across the age groups, particularly for participants in the age group of 25-34, while on the weekend they were more heterogeneous for the participants in the age group of 18-24. Contacts in health locations were mostly with older contacts for all the age groups both on the weekdays and the weekends. Expectedly, homophilic contact patterns were more prevalent for leisure activities on both days, while contact patterns were more heterogeneous on public transport, especially on weekday.

\subsubsection*{Association of socioeconomic and demographic factors with contact patterns}
\begin{figure}[!htb]
  \centering
  \includegraphics[width=1\linewidth]{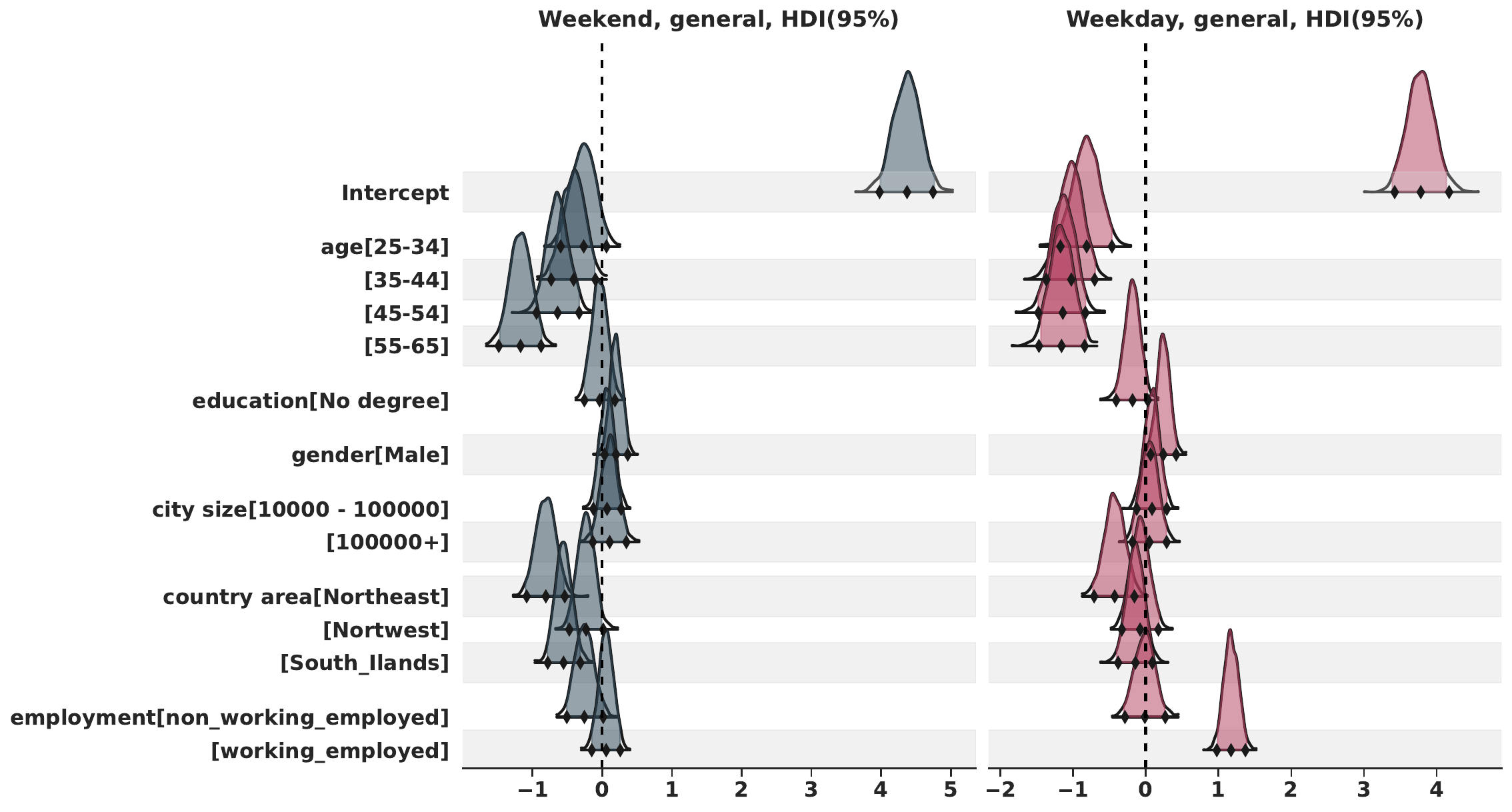}
  \caption{Highest Density Interval (HDI) of the posterior distribution for the negative binomial model of the number of individual contacts per participant on weekend and weekday. On the y-axis the names of each category of the explanatory variables except for the ones included in the intercept, i.e. age[18-24], education[degree], city size[0-10000], country area[Center], employment[unemployed]. On the x-axis the mean number of contacts from the model. The results are to be considered compared to the intercept.}
  \label{fig:cnt_bn_general}
\end{figure}

\begin{figure}[!htb]
  \centering
  \includegraphics[width=1\linewidth]{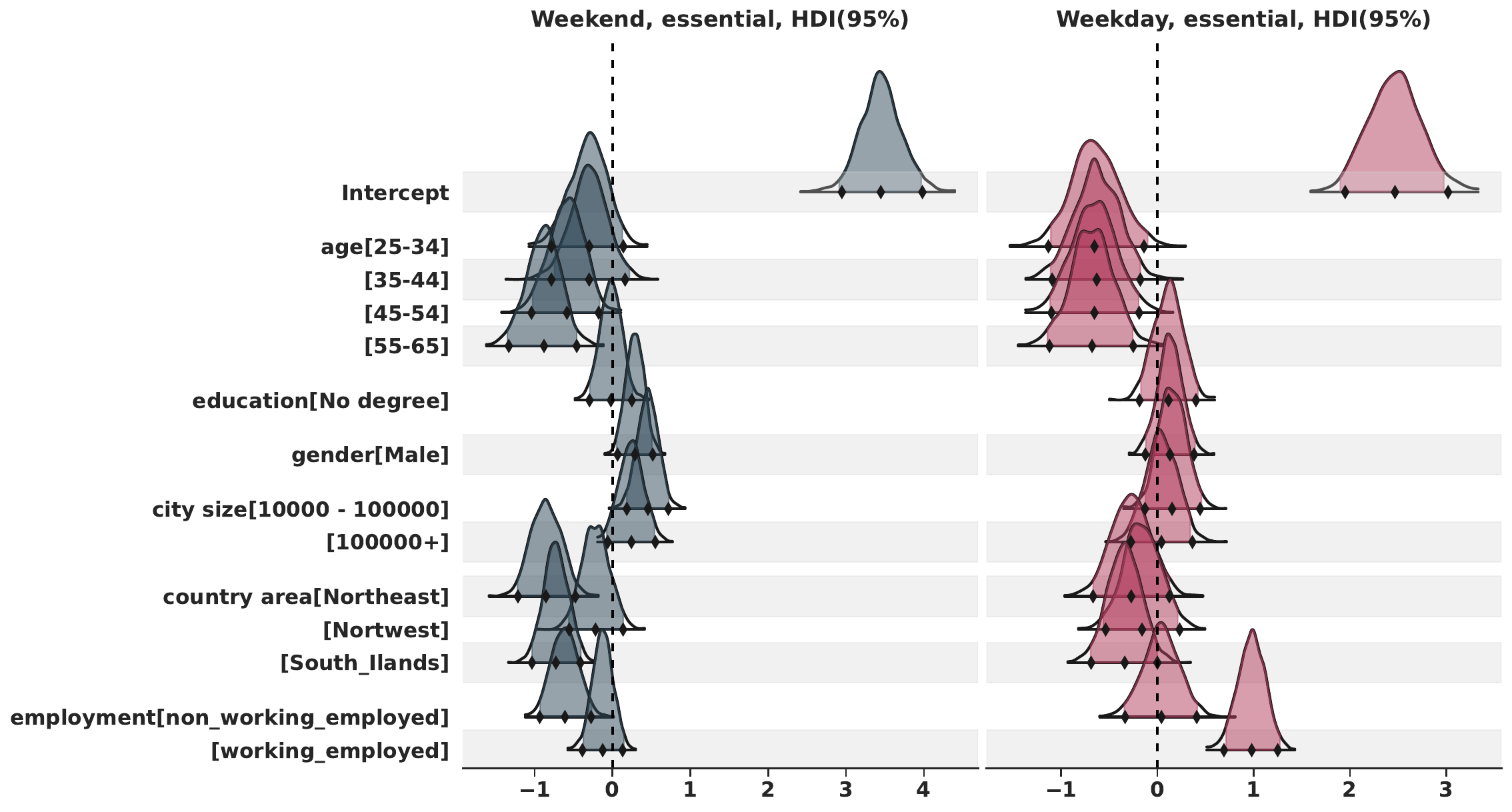}
  \caption{Highest Density Interval (HDI) of the posterior distribution for the negative binomial model during essential activities, on weekend and weekday. On the y-axis the names of each category of the explanatory variables except for the ones included in the intercept, i.e. age[18-24], education[degree], city size['0-10000], country area[Center], employment[unemployed]. On the x-axis the mean number of contacts from the model. The results are to be considered compared to the intercept.}
  \label{fig:cnt_essential}
\end{figure}

\begin{figure}[!hb]
  \centering
  \includegraphics[width=1\linewidth]{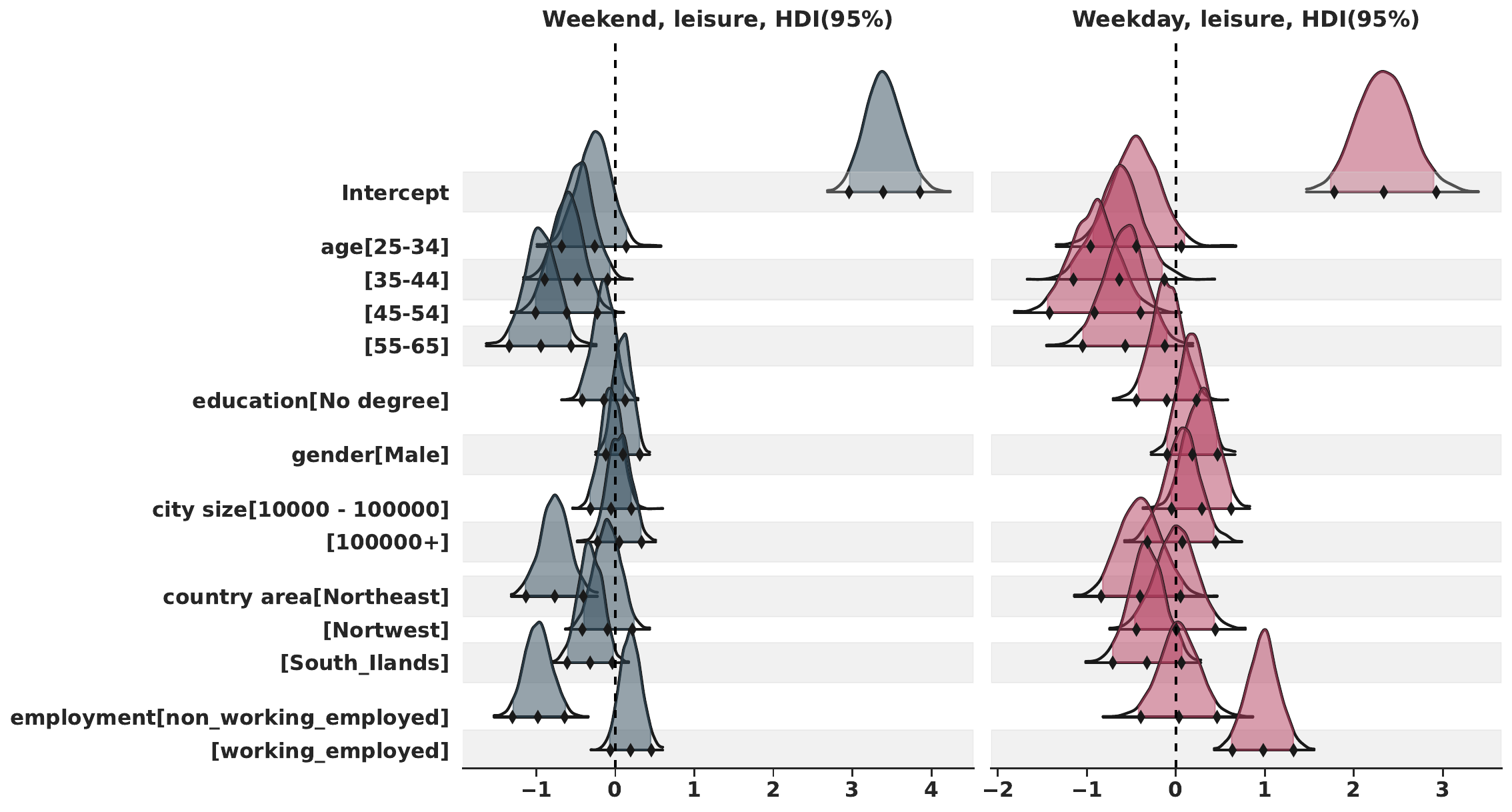}
  \caption{Highest Density Interval (HDI) of the posterior distribution for the negative binomial model during leisure activities, on weekend and weekday. On the y-axis the names of each category of the explanatory variables except for the ones included in the intercept, i.e. age[18-24], education[degree], city size['0-10000], country area[Center], employment[unemployed]. On the x-axis the mean number of contacts from the model. The results are to be considered compared to the intercept.}
  \label{fig:cnt_leisure}
\end{figure}

\begin{figure}[!hb]
  \centering
  \includegraphics[width=1\linewidth]{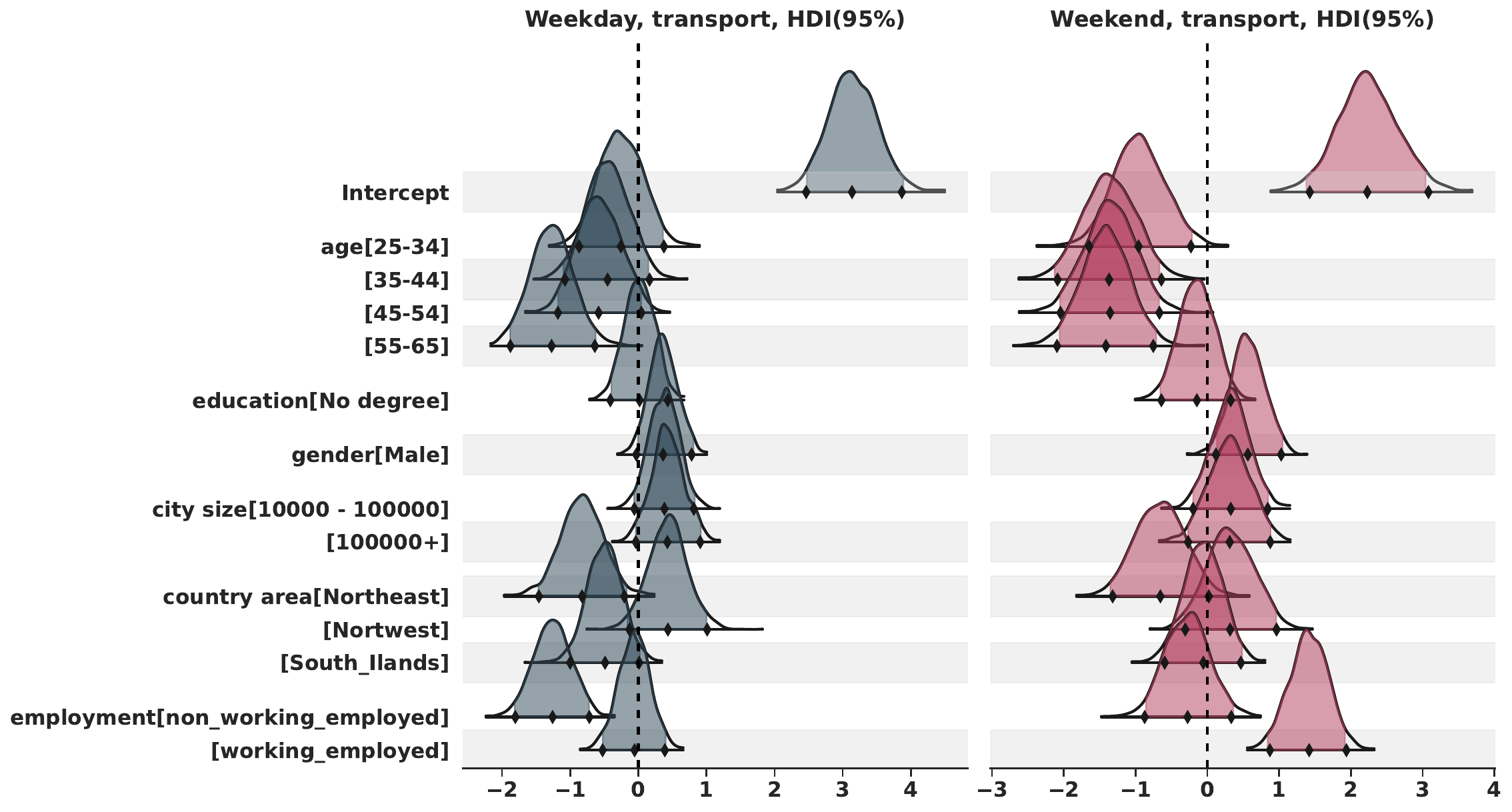}
  \caption{Highest Density Interval (HDI) of the posterior distribution of the negative binomial model on transportation, on weekend and weekday. On the y-axis the names of each category of the explanatory variables except for the ones included in the intercept, i.e. age[18-24], education[degree], city size[0-10000], country area[Center], employment[unemployed]. On the x-axis the mean number of contacts from the model. The results are to be considered compared to the intercept.}
  \label{fig:cnt_transport}
\end{figure}

\begin{figure}[!hb]
  \centering
  \includegraphics[width=1\linewidth]{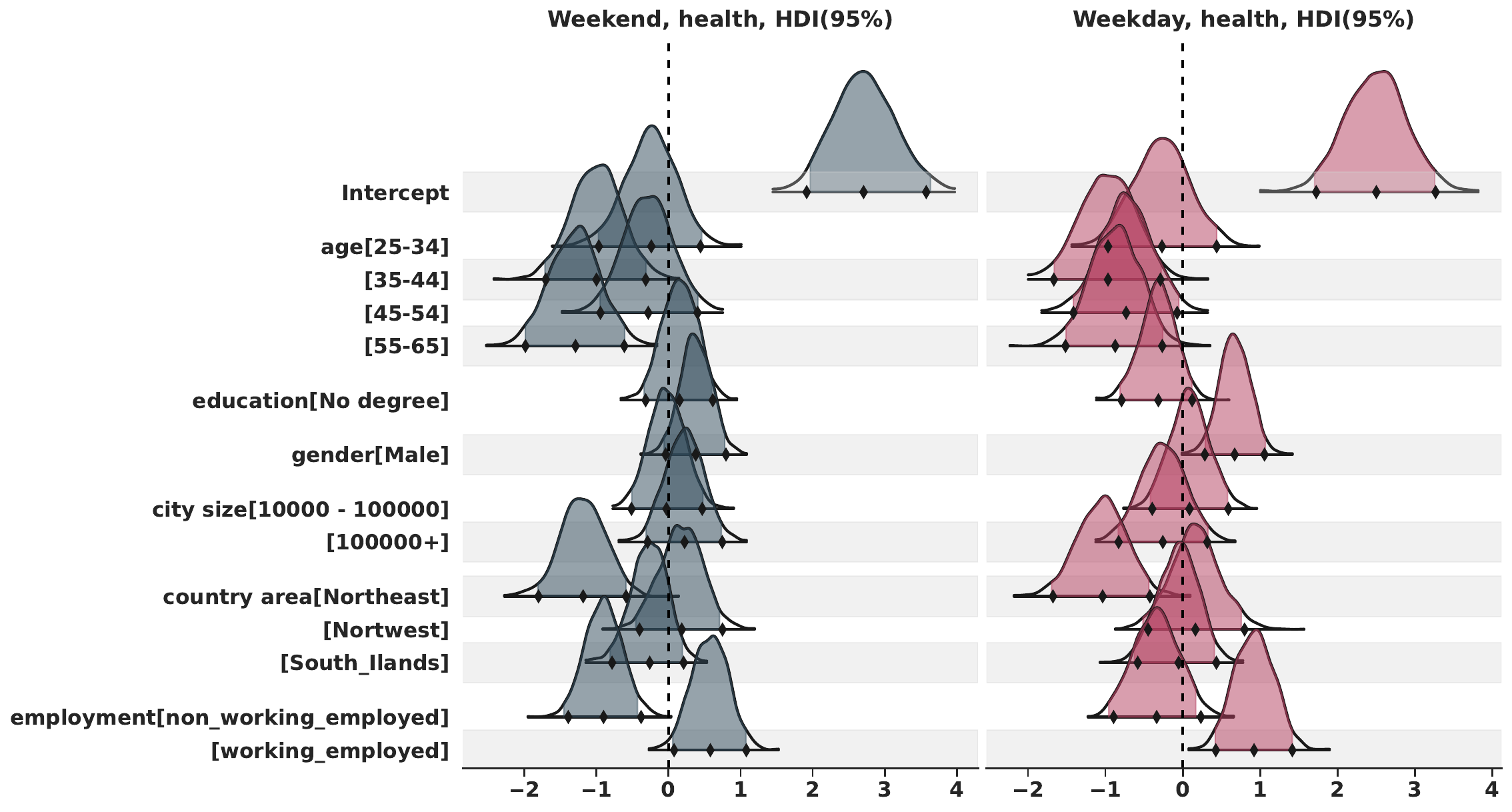}
  \caption{Highest Density Interval (HDI) of the posterior distribution of the negative binomial model in health locations, on weekend and weekday. On the y-axis the names of each category of the explanatory variables except for the ones included in the intercept, i.e. age[18-24], education[degree], city size[0-10000], country area[Center], employment[unemployed]. On the x-axis the mean number of contacts from the model. The results are to be considered compared to the intercept.}
  \label{fig:cnt_health}
\end{figure}

To analyze the distribution of contacts across the two survey days, we explored two models, one for collective and the other for individual contacts. The results of the two models are compatible, so for simplicity, we report only the individual contacts model, while the results of the collective contacts model are in the \nameref{SI_general_nb}. Figure \ref{fig:cnt_bn_general} shows the posterior mean effects and credible intervals for the number of contacts across different strata. Overall, younger participants were more likely to have more contacts for both survey days, tables for the results are shown in the \nameref{SI_general_nb}. Unemployed participants tended to have more contacts on weekend compared to the non-working employed ($mean=-0.26$, $HDI=[-0.51,-0.01]$) and fewer on weekday compared to working employed participants ($mean=0.98$, $HDI=[0.71,1.26]$). Male participants were also more likely to have more contacts than females (weekday: $mean=-0.26$, $HDI=[-0.51,-0.01]$, weekend: $mean=0.2$, $HDI=[0.04,0.35]$). Compared to participants living in the center regions, those in the northeast (weekday: $mean=-0.43$, $HDI=[-0.71,-0.17]$, weekend: $mean=0.81$, $HDI=[-1.05,-0.52]$), south and islands (weekend: $mean=-0.55$, $HDI=[-0.77,-0.32]$), and northwest (weekend: $mean=-0.23$, $HDI=[-0.49,-0.02]$) tended to have fewer contacts. 
To further assess the role of contact location and types of participants' activities in contact behavior, in the following part we employ the model for different places.
\\\textbf{Essential activities.} In Figure \ref{fig:cnt_essential} we show the results of the negative binomial model for contacts during essential activities. Younger participants were overall more likely to have more contacts than older participants, details are shown in the \nameref{SI_essential_nb}. Unemployment was a significant predictor of a higher number of contacts on weekend compared to the non-working employed participants  ($mean=-0.61$, $HDI=[-0.93,-0.29]$). On the other hand, on weekday the tendency was reversed, and unemployed participants were more likely to have fewer contacts than the working participants  ($mean=0.98$, $HDI=[0.71,1.26]$). Additionally, male participants were more likely to have more contacts than female participants on weekend ($mean=0.29$, $HDI=[0.05,0.49]$). Participants living in the northeast were less likely to have contacts than participants living in the central regions on weekend ($mean=-0.85$, $HDI=[-1.23,-0.51]$), likewise for participants living in the south and the islands ($mean=-0.73$, $HDI=[-1.01,-0.4]$). Finally, participants living in medium-sized cities were more likely to have more contact than people living in smaller cities ($mean=0.45$, $HDI=[0.18,0.7]$)
\\\textbf{Leisure activities.} In Figure \ref{fig:cnt_leisure} we show the results for contacts that took place during leisure activities. The number of contacts was driven by the same variables significant for the number of contact during essential activities, except for the city size which in this case was not relevant. Unemployment was a significant predictor of a higher number of contacts on weekend compared to the non-working employed participants  ($mean=-0.98$, $HDI=[-1.29,-0.65]$). On the other hand, on weekday unemployed participants were more likely to have fewer contacts than the working participants ($mean=0.98$, $HDI=[0.64,1.3]$). Participants living in the northeast were less likely to have contacts than participants living in the central regions on weekend ($mean=-0.76$, $HDI=[-1.12,-0.4]$), likewise for participants living in the south and the islands ($mean=-0.32$, $HDI=[-0.58,-0.03]$). More detailed information is reported in the \nameref{SI_leisure_nb}.
\\\textbf{Transport.} Figure \ref{fig:cnt_transport} shows the results for contacts on public transport. In this case, the effect of the strata was more heterogeneous. The effect of the age group was compatible with the previous cases. Unemployed participants were more likely to have contacts compared to the non-working employed on weekend ($mean=-1.26$, $HDI=[-1.81,-0.75]$) while working participants were more likely to have more contacts on weekday ($mean=1.42$, $HDI=[0.85,1.89]$). Participants living in the northeast were less likely to have contacts than participants living in the central regions on weekend ($mean=-0.83$, $HDI=[-1.42,-0.2]$). 
\\\textbf{Health locations.} In Figure \ref{fig:cnt_health} we show the results for contacts in health locations. On weekend unemployed participants were more likely to have more contacts than the non-working employed ($mean=-0.9$, $HDI=[-1.44,-0.46]$), but less than the working participants ($mean=0.58$, $HDI=[0.1,1.07]$). On weekday unemployed participants were more likely to have fewer contacts than working participants ($mean=0.92$, $HDI=[0.46,1.41]$). Male participants were more likely to have more contacts than female, on weekday ($mean=1.42$, $HDI=[0.85,1.89]$). Overall, participants living in northeastern regions were less likely to have contacts compared to those living in the central region (weekday: $mean=-1.04$, $HDI=[-1.65,-0.43]$, weekend: $mean=-1.18$, $HDI=[-1.75,-0.57]$).
\subsubsection*{Socioeconomic determinants of contact behavior}
Overall younger participants had more contacts compared to their counterparts. Geographical differences were also observed, with participants living in the northeastern, southern, and northwestern regions tending to have fewer contacts than those in the central regions.
The analysis of contacts during essential activities, leisure activities, public transport, and health locations displayed similar patterns. Unemployment was a significant predictor of a higher number of contacts on weekends compared to non-working employed participants, but a lower number of contacts on weekdays compared to working participants. Participants living in the northeastern and southern regions consistently had fewer contacts than those in the central regions. Additionally, participants living in medium-sized cities were more likely to have more contact during essential activities compared to those in smaller cities.
These findings highlight the complex interplay between individual characteristics, employment status, and geographic location in shaping contact patterns, which is crucial for understanding the transmission dynamics of communicable diseases.

\section*{Discussion}

In this study, we conducted a survey disaggregated by socioeconomic and demographic strata, offering initial insights into the relationships between these determinants and the behaviors relevant to the spread of infectious diseases in Italy following the COVID-19 pandemic. 
To have a broader overview of the heterogeneous impact of the COVID-19 pandemic we first looked at the association  between the demographic and socioeconomic characteristics of the participants and their perceptions of the  impact on their social, economic, psychological, and occupational situation. 
Age emerged as a key demographic determinant, with younger individuals reporting a more intense experience of the pandemic's effects, supporting the finding in the literature \cite{Pierce2020}. Gender also was found important in the perceived negative effects of the pandemic, particularly concerning psychological well-being, as female participants indicated a higher perceived burden, confirming previous studies \cite{Pierce2020}. Furthermore, the perception of the economic burden was influenced by the participant's city of residence, with those living in larger urban centers reporting a lesser perceived impact on their economic situation. This may be indicative of the differential access to economic resources and opportunities between urban and rural areas, as well as the varying severity of pandemic-related disruptions in different geographical areas \cite{Mindes2024,arin_urban_2022}. \\
In terms of socioeconomic factors, employment status was a persistent determinant affecting all indicators of well-being measured as perception of a negative impact on the participants' situation. Compatible with previous literature \cite{Arena2022, Pierce2020, deGirolamo2022}, unemployed participants reported experiencing a greater negative impact of the pandemic from the psychological point of view, compared to their employed counterparts. Employment status was also systematically associated with a perceived negative impact on economic and occupational well-being, as previously observed during the pandemic  \cite{Vieira2021}. 
Adding to the literature about the heterogeneous impact of the pandemic, our study reveals the heterogeneous perceived impact of the pandemic's burden across different socioeconomic statuses (SES) several years after the start of the pandemic. This underscores the critical importance of including these factors when considering the spread of infectious diseases.

To this aim, we analyzed the association between socioeconomic determinants and both, protective and contact behaviors, key elements in the infectious disease transmission dynamics, while controlling for other known important demographic factors.
Indeed, age was seen as a key determinant of adherence to non-pharmaceutical interventions (NPIs) in the study.
Older participants were more likely to consistently adopt face masks and practice social distancing compared to their younger counterparts. This aligns with existing evidence that risk perception and health-protective behaviors tend to increase with age  \cite{litwin_network-exposure_2021,lages_relation_2021, pasion_age_2020}.  It is important to notice, the survey was administered in March 2024 when both flu and COVID-19 were still circulating among the population.
In terms of contact behavior, younger participants reported higher overall contact levels compared to older age groups. This aligns with previous research indicating that younger individuals tend to have a more rich social network \cite{Dorlien2021, mossong2008social}. Aligned with previous results, male participants also were more likely to have more contacts than females \cite{Bhattacharya2016}. 
These observations corroborate those of previous research concerning the association  between age and both protective and contact behaviors, even in the context of the post-pandemic era.\\

Turning to less explored factors we found socioeconomic characteristics to be key determinants of the behaviors relevant to the spread of diseases. In particular, employment status exhibited an inverse relationship between NPIs and vaccine uptake. Unemployed participants were more likely to consistently wear masks and maintain social distancing, potentially reflecting their heightened concerns about infection risk. However, they were less likely to be vaccinated against the virus. This underscores the multifaceted nature of vaccine hesitancy, which can arise from various socioeconomic barriers as seen in previous literature \cite{rhodes_potential_2024, beale_occupation_2022}. 

On the contact behavior side, our analysis shows the importance of employment status in shaping contact patterns as previously observed in \cite{manna_importance_2024} during the pandemic period. This shows that there seems to be a persistent association between employment and contact behavior long after the end of the pandemic. Moreover, we uncover an additional layer of complexity, revealing a non-trivial relationship between employment status, weekdays, and weekends in relation to contact patterns. Working-employed participants were more likely to have higher contact levels on weekdays compared to unemployed individuals, who, conversely, reported more contacts on weekends than their non-working employed counterparts. This finding suggests that employment status not only affects overall contact behavior but also interacts with the temporal aspect of contacts, with weekday and weekend contact patterns differing based on employment status. \\
Overall, these findings have significant implications for understanding the drivers of social patterns that underpin disease transmission. In particular, the recognition of the socioeconomic determinants of contact patterns will assist in the design of more effective epidemic models of communicable diseases. Pragmatically, employment status emerged as a persistent factor shaping contact behavior, mask-wearing, and social distancing, instead, education was found to be linked to the influenza vaccination uptake. 
While the survey and its analysis are an important step to assess the interplay between socioeconomic factors and, contact and protective behaviors, it faces limitations due to its defined age range considered by the panel, which did not include participants younger than 18 years old, and older than 65. Both of these age groups possess distinct epidemiological relevance and may exhibit different behavioral patterns, particularly in the underage population.  The second limitation was the use of a computer-assisted interview, which might create a biased sample of the population toward higher socioeconomic classes. Despite these limitations, our findings reveal significant heterogeneity in the associations between socioeconomic factors and behaviors relevant the spread of a disease. This underscores the necessity of considering socioeconomic factors when integrating human behavior in epidemic modeling.
Finally, the survey's representation of independent population strata complicates the interpretation of stratified analyses, particularly in constructing contact matrices, highlighting a methodological challenge for small sample-sized surveys that could benefit from further research.

\section*{Acknowledgment}
This project has received funding from the Italian Fund for Economic and Social Research (FRES2021\_0000002).
L.G. and M. T. acknowledge the support from the Lagrange Project of the Institute for Scientific Interchange Foundation (ISI Foundation) funded by the Fondazione Cassa di Risparmio di Torino (Fondazione CRT).
Founders did not play any role in the study design.

\clearpage

\section*{Supporting Information}
\subsection*{Survey}
\label{SI.0}

\textbf{A1} Gender:
\begin{enumerate}
\item Male
\item Female
\end{enumerate}

\textbf{A2} Age:

\textbf{A3} Indicate your municipality of residence:
(Display list of municipalities)
CED: Generate province, region, and size of the municipality from Istat database.
Recode municipality size into:
\begin{enumerate}
\item Less than 10,000 inhabitants
\item Between 10,000 and 100,000 inhabitants
\item Over 100,000 inhabitants
\end{enumerate}

\textbf{A5} What is the highest educational qualification you have obtained?
\begin{enumerate}
\item Post-graduate degree (PhD/Master's)
\item University degree (5-year program or equivalent)
\item University degree (3-year program)
\item High school diploma
\item Middle school certificate
\item Elementary school certificate
\item No qualification
\end{enumerate}

\textbf{A6} Indicate your current occupation/employment status:
\begin{itemize}
    \item \textit{EMPLOYED}
    \item \textit{UNEMPLOYED }
\end{itemize}

\textbf{D1} In the last 6 months, have you had the flu vaccine?
\begin{enumerate}
\item Yes
\item No
\item Prefer not to answer
\end{enumerate}

\textbf{D2} Have you had the Covid-19 vaccine? If so, how many doses have you had?
\begin{enumerate}
\item 1
\item 2
\item 3
\item More than 3
\item No, none
\item Prefer not to answer
\end{enumerate}

\textbf{D3} In the last month, have you used the following preventive measures against infectious diseases (Covid-19 and/or flu)?
Always - Sometimes - Never
\begin{itemize}
\item Protective mask
\item Social distancing
\end{itemize}

\textbf{D4} Now read some statements about the impact of Covid-19 on various aspects of life. Thinking about your personal situation, indicate how much you agree with each statement using a scale from 1 to 5, where 1 means "not at all agree" and 5 means "strongly agree".

\begin{enumerate}
\item Strongly agree
\item Somewhat agree
\item Neither agree nor disagree
\item Slightly disagree
\item Not at all agree
\end{enumerate}
\begin{itemize}
\item Covid has had a negative impact on my economic situation
\item Covid has had a negative impact on my social situation (e.g. I've lost contact with friends, I participate less in social events, etc.)
\item Covid has had a negative impact on my psychological condition (e.g. I feel more mentally fatigued, I feel more stressed...)
\item Covid has had a negative impact on my work situation (e.g. I lost my job, it was more difficult to find employment)
\end{itemize}

\textbf{D8} Did you work last Sunday?
\begin{enumerate}
\item Yes
\item No
\end{enumerate}

\textbf{D9} Think about the people you came into contact with last Sunday, between 00:00 and 24:00. We are interested only in direct contacts, that is, people you met in person and with whom you exchanged at least a few words or had physical contact (e.g. a handshake, a hug, a kiss, a contact sport...). Exclude people you interacted with only by phone or online. We will propose some meeting places to you below. For each of these, indicate, for each age group, how many people you have had direct contact with. Set the minimum response number to 0 and the maximum to 100 for each item.

We will propose some meeting places to you below. For each of these, indicate, for each age group, how many people you have had direct contact with.
Set the minimum response number to 0 and the maximum to 100 for each item.

\textbf{Inside your own home}
(excluding members of your own family):

\textbf{Number of people:}
\begin{itemize}
    \item Age between 0 and 5 years
    \item Age between 6 and 17 years
    \item Age between 18 and 24 years
    \item Age between 25 and 35 years
    \item Age between 36 and 45 years
    \item Age between 46 and 55 years
    \item Age between 56 and 65 years
    \item Age over 65 years
\end{itemize}

\textbf{If code 1 to D8 (worked last Sunday) or code 17 to A6 (student)
In workplaces or study (work/ school / university):}

\textbf{Number of people:}
\begin{itemize}
    \item Age between 0 and 5 years
    \item Age between 6 and 17 years
    \item Age between 18 and 24 years
    \item Age between 25 and 35 years
    \item Age between 36 and 45 years
    \item Age between 46 and 55 years
    \item Age between 56 and 65 years
    \item Age over 65 years
\end{itemize}

\textbf{On public transport:}

\textbf{Number of people:}
\begin{itemize}
    \item Age between 0 and 5 years
    \item Age between 6 and 17 years
    \item Age between 18 and 24 years
    \item Age between 25 and 35 years
    \item Age between 36 and 45 years
    \item Age between 46 and 55 years
    \item Age between 56 and 65 years
    \item Age over 65 years
\end{itemize}

\textbf{Performing essential activities (e.g., grocery shopping, etc.):}

\textbf{Number of people:}
\begin{itemize}
    \item Age between 0 and 5 years
    \item Age between 6 and 17 years
    \item Age between 18 and 24 years
    \item Age between 25 and 35 years
    \item Age between 36 and 45 years
    \item Age between 46 and 55 years
    \item Age between 56 and 65 years
    \item Age over 65 years
\end{itemize}

For all
Performing health-related activities
(medical visits, hospitals, laboratory tests):

\textbf{Number of people:}
\begin{itemize}
    \item Age between 0 and 5 years
    \item Age between 6 and 17 years
    \item Age between 18 and 24 years
    \item Age between 25 and 35 years
    \item Age between 36 and 45 years
    \item Age between 46 and 55 years
    \item Age between 56 and 65 years
    \item Age over 65 years
\end{itemize}

\textbf{Performing leisure activities (sports, concerts, events, restaurants, etc.):}

\textbf{Number of people:}
\begin{itemize}
    \item Age between 0 and 5 years
    \item Age between 6 and 17 years
    \item Age between 18 and 24 years
    \item Age between 25 and 35 years
    \item Age between 36 and 45 years
    \item Age between 46 and 55 years
    \item Age between 56 and 65 years
    \item Age over 65 years
\end{itemize}

\textbf{D10} Still thinking about Sunday, between 00:00 and 24:00, please indicate how many people you had indirect contacts with, that is, people you spent more than 15 minutes with in the same place (e.g. on public transport...) and with whom you maintained a distance of 1.5 meters or less. Exclude the direct contacts already reported in the previous question and people you interacted with only by phone or online.

As above, we will propose some meeting places below. For each of these, indicate how many people you have had indirect contact with during the day.
Set the minimum response number to 0 and the maximum to 100 for each item.

\begin{itemize}
    \item Inside your own home (excluding members of your own family):
\end{itemize}
Number of people:

If code 1 to D8 (worked last Sunday) or code 17 to A6 (student)
\begin{itemize}
    \item In workplaces or study (work/school/university):
\end{itemize}

Number of people:

\begin{itemize}
    \item On public transport:
\end{itemize}

Number of people:

\begin{itemize}
    \item Performing essential activities (e.g., shopping, etc.):
\end{itemize}

Number of people:

\begin{itemize}
    \item Performing health-related activities (medical visits, hospitals, laboratory tests):
\end{itemize}

Number of people:

\begin{itemize}
    \item Performing leisure activities (sports, concerts, events, restaurants, etc.):
\end{itemize}

Number of people:
\\
\textbf{D11} Did you work last weekday?
\begin{enumerate}
\item Yes
\item No
\end{enumerate}

\textbf{D12} Now think about Monday, between 00:00 and 24:00. How many people did you come into contact with? We are interested only in direct contacts, that is, people you met in person and with whom you exchanged at least a few words or had physical contact (e.g. a handshake, a hug, a kiss, a contact sport...). Exclude people you interacted with only by phone or online. We will propose some meeting places to you below. For each of these, indicate, for each age group, how many people you have had direct contact with.
Set the minimum response number to 0 and the maximum to 100 for each item.

\textbf{Inside your own home}
(excluding members of your own family):

\textbf{Number of people:}
\begin{itemize}
    \item Age between 0 and 5 years
    \item Age between 6 and 17 years
    \item Age between 18 and 24 years
    \item Age between 25 and 35 years
    \item Age between 36 and 45 years
    \item Age between 46 and 55 years
    \item Age between 56 and 65 years
    \item Age over 65 years
\end{itemize}

\textbf{If code 1 to D8 (worked last Sunday) or code 17 to A6 (student)
In workplaces or study (work/ school / university):}

\textbf{Number of people:}
\begin{itemize}
    \item Age between 0 and 5 years
    \item Age between 6 and 17 years
    \item Age between 18 and 24 years
    \item Age between 25 and 35 years
    \item Age between 36 and 45 years
    \item Age between 46 and 55 years
    \item Age between 56 and 65 years
    \item Age over 65 years
\end{itemize}

\textbf{On public transport:}

\textbf{Number of people:}
\begin{itemize}
    \item Age between 0 and 5 years
    \item Age between 6 and 17 years
    \item Age between 18 and 24 years
    \item Age between 25 and 35 years
    \item Age between 36 and 45 years
    \item Age between 46 and 55 years
    \item Age between 56 and 65 years
    \item Age over 65 years
\end{itemize}

\textbf{Performing essential activities (e.g., grocery shopping, etc.):}

\textbf{Number of people:}
\begin{itemize}
    \item Age between 0 and 5 years
    \item Age between 6 and 17 years
    \item Age between 18 and 24 years
    \item Age between 25 and 35 years
    \item Age between 36 and 45 years
    \item Age between 46 and 55 years
    \item Age between 56 and 65 years
    \item Age over 65 years
\end{itemize}

For all
Performing health-related activities
(medical visits, hospitals, laboratory tests):

\textbf{Number of people:}
\begin{itemize}
    \item Age between 0 and 5 years
    \item Age between 6 and 17 years
    \item Age between 18 and 24 years
    \item Age between 25 and 35 years
    \item Age between 36 and 45 years
    \item Age between 46 and 55 years
    \item Age between 56 and 65 years
    \item Age over 65 years
\end{itemize}

\textbf{Performing leisure activities (sports, concerts, events, restaurants, etc.):}

\textbf{Number of people:}
\begin{itemize}
    \item Age between 0 and 5 years
    \item Age between 6 and 17 years
    \item Age between 18 and 24 years
    \item Age between 25 and 35 years
    \item Age between 36 and 45 years
    \item Age between 46 and 55 years
    \item Age between 56 and 65 years
    \item Age over 65 years
\end{itemize}

\textbf{D10} Still thinking about Monday, between 00:00 and 24:00, please indicate how many people you had indirect contacts with, that is, people you spent more than 15 minutes with in the same place (e.g. on public transport...) and with whom you maintained a distance of 1.5 meters or less. Exclude the direct contacts already reported in the previous question and people you interacted with only by phone or online. As above, we will propose some meeting places below. For each of these, indicate how many people you have had indirect contact with during the day.
Set the minimum response number to 0 and the maximum to 100 for each item.

\begin{itemize}
    \item Inside your own home (excluding members of your own family):
\end{itemize}
Number of people:

If code 1 to D8 (worked last Sunday) or code 17 to A6 (student)
\begin{itemize}
    \item In workplaces or study (work/school/university):
\end{itemize}

Number of people:

\begin{itemize}
    \item On public transport:
\end{itemize}

Number of people:

\begin{itemize}
    \item Performing essential activities (e.g., shopping, etc.):
\end{itemize}

Number of people:

\begin{itemize}
    \item Performing health-related activities (medical visits, hospitals, laboratory tests):
\end{itemize}

Number of people:

\begin{itemize}
    \item Performing leisure activities (sports, concerts, events, restaurants, etc.):
\end{itemize}

Number of people:
\\
\subsection*{Survey descriptive statistics}

\begin{table}[H]
\let\center\empty
\let\endcenter\relax
\centering
\resizebox{.9\width}{!}{\begin{tabular}{|l|l|p{2cm}|p{2cm}|p{2cm}|}
\hline
               &    &  Number of Participants &  Percentage (\%) & National distribution (\%)\\
\hline
Age group & 18 – 24 &    132 &       11.00 & 11.2 \\
               & 25 – 34 &    205 &       17.08 & 17.1 \\
               & 35 – 44 &    237 &       19.75  & 19.5 \\
               & 45 – 54 &    306 &       25.50 & 25.5 \\
               & 55 – 65 &    320 &       26.67  & 26.7 \\ \hline
Gender & Woman &    599 &       49.92 & 49.9 \\
               & Man &    601 &       50.08 & 50.1 \\ \hline
Geographical Area & Center &    237 &       19.75 & 19.8 \\
               & Northeast &    234 &       19.50  & 19.5 \\
               & Northwest &    318 &       26.50  & 26.7 \\
               & South and Islands &    411 &       34.25 & 34\\ \hline
Center's Size & Up to 10.000 &    377 &       31.42 & 30.4 \\
                & 10.000 - 100.000 &    540 &       45.00 & 46.5 \\
                & More than 100.000 &    283 &       23.58 & 23.1 \\ \hline
Employment status & Not working & 454 & 37.8  & 40.1\\
                  & Working     & 740 &  61.7 & 59.9\\
                  & No answer   &  6  &  0.5 & \_  \\ \hline
Education   & With degree & 252 & 21 & 20.4 \\
            & Without degree & 948 & 79 & 79.6 \\ \hline 

\end{tabular}
}
\label{tabsi:descr_stat}
    \supptable{Descriptive statistics}
\end{table}

\subsection*{Results of the Bayesian negative binomial model for individual and collective contacts}

To estimate the Bayesian negative binomial regression model, we used Markov Chain Monte Carlo (MCMC) sampling techniques \cite{sammut_markov_2010}. We specified non-informative prior distributions for the model parameters and ran multiple MCMC chains to ensure convergence. The MCMC samples were used to obtain posterior estimates of the model parameters and to quantify the uncertainty in these estimates.
These diagnostic tests and model comparison methods ensured that our models met the necessary assumptions and provided a good fit to our data, allowing us to draw reliable conclusions from the results using Bayesian inference.\\

\subsubsection*{General settings}

\paragraph*{SI1.1}
\label{SI_general_nb}
\textbf{Monday individual model}


\begin{table}[H]
\let\center\empty
\let\endcenter\relax
\centering
\resizebox{.9\width}{!}{\begin{tabular}{lllll}
\toprule
 & mean & sd & hdi\_3\% & hdi\_97\% \\
\midrule
Intercept & 3.78 & 0.19 & 3.4 & 4.12 \\
age\_group[25-34] & -0.81 & 0.18 & -1.16 & -0.47 \\
age\_group[35-44] & -1.03 & 0.17 & -1.36 & -0.72 \\
age\_group[45-54] & -1.14 & 0.17 & -1.46 & -0.84 \\
age\_group[55-65] & -1.16 & 0.16 & -1.44 & -0.84 \\
city\_size[10000 - 100000] & 0.08 & 0.11 & -0.13 & 0.27 \\
city\_size[100000+] & 0.05 & 0.12 & -0.18 & 0.27 \\
cnt\_id\_alpha & 0.47 & 0.02 & 0.43 & 0.5 \\
country\_area[Northeast] & -0.43 & 0.14 & -0.71 & -0.17 \\
country\_area[Nortwest] & -0.08 & 0.13 & -0.31 & 0.18 \\
country\_area[South\_Ilands] & -0.14 & 0.12 & -0.37 & 0.08 \\
education[Nodegree] & -0.18 & 0.11 & -0.39 & 0.03 \\
employment[non\_working\_employed] & -0.01 & 0.14 & -0.3 & 0.23 \\
employment[working\_employed] & 1.17 & 0.1 & 0.99 & 1.37 \\
gender[Male] & 0.24 & 0.09 & 0.07 & 0.41 \\
\bottomrule
\end{tabular}

}
\supptable{Results of the Negative Binomial Model for individual contacts on Monday without a specific location}
\label{tabsi:mon_all_ind}
\end{table}
\textbf{Monday collective model}


\begin{table}[H]
\let\center\empty
\let\endcenter\relax
\centering
\resizebox{.9\width}{!}{\begin{tabular}{lllll}
\toprule
 & mean & sd & hdi\_3\% & hdi\_97\% \\
\midrule
Intercept & 4.17 & 0.19 & 3.83 & 4.52 \\
age\_group[25-34] & -0.9 & 0.18 & -1.23 & -0.56 \\
age\_group[35-44] & -1.03 & 0.17 & -1.34 & -0.71 \\
age\_group[45-54] & -1.14 & 0.17 & -1.47 & -0.83 \\
age\_group[55-65] & -1.2 & 0.16 & -1.52 & -0.94 \\
city\_size[10000 - 100000] & 0.06 & 0.1 & -0.13 & 0.24 \\
city\_size[100000+] & 0.02 & 0.12 & -0.19 & 0.25 \\
cnt\_id\_alpha & 0.47 & 0.02 & 0.44 & 0.5 \\
country\_area[Northeast] & -0.36 & 0.14 & -0.65 & -0.13 \\
country\_area[Nortwest] & 0.01 & 0.13 & -0.25 & 0.25 \\
country\_area[South\_Ilands] & -0.08 & 0.12 & -0.3 & 0.17 \\
education[Nodegree] & -0.19 & 0.1 & -0.37 & 0.01 \\
employment[non\_working\_employed] & -0.09 & 0.14 & -0.36 & 0.17 \\
employment[working\_employed] & 1.15 & 0.1 & 0.94 & 1.33 \\
gender[Male] & 0.27 & 0.09 & 0.12 & 0.44 \\
\bottomrule
\end{tabular}

}
\supptable{Results of the Negative Binomial Model for both individual and collective contacts on Monday without a specific location}
\label{tabsi:mon_all_col}
\end{table}
\textbf{Sunday individual model}


\begin{table}[H]
\let\center\empty
\let\endcenter\relax
\centering
\resizebox{.9\width}{!}{\begin{tabular}{lllll}
\toprule
 & mean & sd & hdi\_3\% & hdi\_97\% \\
\midrule
Intercept & 4.37 & 0.2 & 4.02 & 4.76 \\
age\_group[25-34] & -0.27 & 0.17 & -0.59 & 0.04 \\
age\_group[35-44] & -0.41 & 0.16 & -0.74 & -0.12 \\
age\_group[45-54] & -0.64 & 0.16 & -0.92 & -0.32 \\
age\_group[55-65] & -1.17 & 0.16 & -1.46 & -0.87 \\
city\_size[10000 - 100000] & 0.07 & 0.1 & -0.13 & 0.25 \\
city\_size[100000+] & 0.1 & 0.12 & -0.13 & 0.34 \\
cnt\_id\_alpha & 0.47 & 0.02 & 0.43 & 0.5 \\
country\_area[Northeast] & -0.81 & 0.14 & -1.05 & -0.52 \\
country\_area[Nortwest] & -0.23 & 0.13 & -0.49 & -0.02 \\
country\_area[South\_Ilands] & -0.55 & 0.12 & -0.77 & -0.32 \\
education[Nodegree] & -0.04 & 0.12 & -0.24 & 0.19 \\
employment[non\_working\_employed] & -0.26 & 0.14 & -0.51 & -0.01 \\
employment[working\_employed] & 0.06 & 0.11 & -0.14 & 0.26 \\
gender[Male] & 0.2 & 0.09 & 0.04 & 0.35 \\
\bottomrule
\end{tabular}

}
\supptable{Results of the Negative Binomial Model for individual contacts on weekend without a specific location}
\label{tabsi:sun_all_ind}
\end{table}
\textbf{weekend collective model}


\begin{table}[H]
\let\center\empty
\let\endcenter\relax
\centering
\resizebox{.9\width}{!}{\begin{tabular}{lllll}
\toprule
 & mean & sd & hdi\_3\% & hdi\_97\% \\
\midrule
Intercept & 4.66 & 0.19 & 4.27 & 4.99 \\
age\_group[25-34] & -0.37 & 0.16 & -0.66 & -0.06 \\
age\_group[35-44] & -0.45 & 0.16 & -0.74 & -0.16 \\
age\_group[45-54] & -0.69 & 0.15 & -0.97 & -0.41 \\
age\_group[55-65] & -1.15 & 0.15 & -1.43 & -0.86 \\
city\_size[10000 - 100000] & 0.11 & 0.1 & -0.08 & 0.29 \\
city\_size[100000+] & 0.12 & 0.12 & -0.1 & 0.33 \\
cnt\_id\_alpha & 0.48 & 0.02 & 0.44 & 0.51 \\
country\_area[Northeast] & -0.62 & 0.14 & -0.9 & -0.37 \\
country\_area[Nortwest] & -0.18 & 0.13 & -0.4 & 0.09 \\
country\_area[South\_Ilands] & -0.44 & 0.13 & -0.68 & -0.21 \\
education[Nodegree] & -0.03 & 0.11 & -0.24 & 0.18 \\
employment[non\_working\_employed] & -0.25 & 0.13 & -0.48 & -0.0 \\
employment[working\_employed] & 0.1 & 0.1 & -0.08 & 0.28 \\
gender[Male] & 0.16 & 0.09 & 0.01 & 0.33 \\
\bottomrule
\end{tabular}

}
\supptable{Results of the Negative Binomial Model for both individual and collective contacts on weekend without a specific location}
\label{tabsi:sun_all_col}
\end{table}

\subsubsection*{Essential activities}
\paragraph*{S1.2}
\label{SI_essential_nb}
\textbf{Monday individual}


\begin{table}[H]
\let\center\empty
\let\endcenter\relax
\centering
\resizebox{.9\width}{!}{\begin{tabular}{lllll}
\toprule
 & mean & sd & hdi\_3\% & hdi\_97\% \\
\midrule
Intercept & 2.47 & 0.28 & 1.94 & 2.97 \\
age\_group[25-34] & -0.64 & 0.26 & -1.12 & -0.15 \\
age\_group[35-44] & -0.63 & 0.24 & -1.07 & -0.18 \\
age\_group[45-54] & -0.65 & 0.24 & -1.11 & -0.23 \\
age\_group[55-65] & -0.68 & 0.22 & -1.09 & -0.24 \\
city\_size[10000 - 100000] & 0.16 & 0.15 & -0.1 & 0.46 \\
city\_size[100000+] & 0.05 & 0.17 & -0.27 & 0.34 \\
cnt\_id\_alpha & 0.41 & 0.02 & 0.36 & 0.45 \\
country\_area[Northeast] & -0.27 & 0.21 & -0.64 & 0.13 \\
country\_area[Nortwest] & -0.16 & 0.2 & -0.55 & 0.19 \\
country\_area[South\_Ilands] & -0.34 & 0.18 & -0.69 & -0.02 \\
education[Nodegree] & 0.11 & 0.15 & -0.16 & 0.4 \\
employment[non\_working\_employed] & 0.04 & 0.19 & -0.32 & 0.41 \\
employment[working\_employed] & 0.98 & 0.14 & 0.71 & 1.26 \\
gender[Male] & 0.13 & 0.13 & -0.11 & 0.39 \\
\bottomrule
\end{tabular}

}
\supptable{Results of the Negative Binomial Model for individual contacts on weekday during essential activities}
\label{tabsi:mon_ess_ind}
\end{table}
\textbf{Monday collective}


\begin{table}[H]
\let\center\empty
\let\endcenter\relax
\centering
\resizebox{.9\width}{!}{\begin{tabular}{lllll}
\toprule
 & mean & sd & hdi\_3\% & hdi\_97\% \\
\midrule
Intercept & 2.8 & 0.25 & 2.33 & 3.28 \\
age\_group[25-34] & -0.74 & 0.24 & -1.17 & -0.3 \\
age\_group[35-44] & -0.59 & 0.22 & -1.01 & -0.19 \\
age\_group[45-54] & -0.7 & 0.22 & -1.12 & -0.27 \\
age\_group[55-65] & -0.65 & 0.22 & -1.05 & -0.26 \\
city\_size[10000 - 100000] & 0.17 & 0.14 & -0.08 & 0.43 \\
city\_size[100000+] & 0.08 & 0.16 & -0.2 & 0.37 \\
cnt\_id\_alpha & 0.42 & 0.02 & 0.38 & 0.46 \\
country\_area[Northeast] & -0.23 & 0.19 & -0.57 & 0.11 \\
country\_area[Nortwest] & -0.09 & 0.18 & -0.45 & 0.23 \\
country\_area[South\_Ilands] & -0.29 & 0.16 & -0.6 & -0.0 \\
education[Nodegree] & 0.06 & 0.14 & -0.19 & 0.35 \\
employment[non\_working\_employed] & 0.04 & 0.18 & -0.33 & 0.37 \\
employment[working\_employed] & 0.85 & 0.14 & 0.62 & 1.11 \\
gender[Male] & 0.2 & 0.12 & -0.01 & 0.43 \\
\bottomrule
\end{tabular}

}
\supptable{Results of the Negative Binomial Model for both individual and collective contacts on weekday during essential activities}
\label{tabsi:mon_ess_col}
\end{table}

\textbf{weekend individual}


\begin{table}[H]
\let\center\empty
\let\endcenter\relax
\centering
\resizebox{.9\width}{!}{\begin{tabular}{lllll}
\toprule
 & mean & sd & hdi\_3\% & hdi\_97\% \\
\midrule
Intercept & 3.45 & 0.27 & 2.96 & 3.97 \\
age\_group[25-34] & -0.31 & 0.24 & -0.76 & 0.13 \\
age\_group[35-44] & -0.3 & 0.24 & -0.74 & 0.17 \\
age\_group[45-54] & -0.59 & 0.23 & -1.02 & -0.19 \\
age\_group[55-65] & -0.88 & 0.22 & -1.32 & -0.47 \\
city\_size[10000 - 100000] & 0.45 & 0.14 & 0.18 & 0.7 \\
city\_size[100000+] & 0.24 & 0.16 & -0.05 & 0.54 \\
cnt\_id\_alpha & 0.45 & 0.02 & 0.41 & 0.49 \\
country\_area[Northeast] & -0.85 & 0.2 & -1.23 & -0.51 \\
country\_area[Nortwest] & -0.22 & 0.18 & -0.55 & 0.13 \\
country\_area[South\_Ilands] & -0.73 & 0.16 & -1.01 & -0.4 \\
education[Nodegree] & -0.02 & 0.14 & -0.3 & 0.23 \\
employment[non\_working\_employed] & -0.61 & 0.17 & -0.93 & -0.29 \\
employment[working\_employed] & -0.13 & 0.13 & -0.37 & 0.13 \\
gender[Male] & 0.29 & 0.12 & 0.05 & 0.49 \\
\bottomrule
\end{tabular}

}
\supptable{Results of the Negative Binomial Model for individual contacts on weekend during essential activities}
\label{tabsi:sun_ess_ind}
\end{table}
\textbf{weekend collective}


\begin{table}[H]
\let\center\empty
\let\endcenter\relax
\centering
\resizebox{.9\width}{!}{\begin{tabular}{lllll}
\toprule
 & mean & sd & hdi\_3\% & hdi\_97\% \\
\midrule
Intercept & 3.78 & 0.24 & 3.34 & 4.22 \\
age\_group[25-34] & -0.39 & 0.22 & -0.8 & 0.03 \\
age\_group[35-44] & -0.34 & 0.22 & -0.74 & 0.07 \\
age\_group[45-54] & -0.46 & 0.2 & -0.82 & -0.07 \\
age\_group[55-65] & -0.78 & 0.2 & -1.17 & -0.42 \\
city\_size[10000 - 100000] & 0.3 & 0.12 & 0.07 & 0.52 \\
city\_size[100000+] & 0.11 & 0.15 & -0.17 & 0.4 \\
cnt\_id\_alpha & 0.49 & 0.02 & 0.44 & 0.53 \\
country\_area[Northeast] & -0.7 & 0.17 & -1.02 & -0.38 \\
country\_area[Nortwest] & -0.18 & 0.16 & -0.49 & 0.12 \\
country\_area[South\_Ilands] & -0.66 & 0.15 & -0.94 & -0.4 \\
education[Nodegree] & -0.04 & 0.14 & -0.3 & 0.2 \\
employment[non\_working\_employed] & -0.58 & 0.16 & -0.88 & -0.27 \\
employment[working\_employed] & -0.07 & 0.13 & -0.3 & 0.17 \\
gender[Male] & 0.22 & 0.11 & -0.0 & 0.42 \\
\bottomrule
\end{tabular}

}
\supptable{Results of the Negative Binomial Model for both individual and collective contacts on weekend during essential activities}
\label{tabsi:sun_ess_all}
\end{table}

\subsubsection*{Leisure activities}
\paragraph*{S1.3}
\label{SI_leisure_nb}
\textbf{weekday individual}


\begin{table}[H]
\let\center\empty
\let\endcenter\relax
\centering
\resizebox{.9\width}{!}{\begin{tabular}{lllll}
\toprule
 & mean & sd & hdi\_3\% & hdi\_97\% \\
\midrule
Intercept & 2.34 & 0.3 & 1.78 & 2.9 \\
age\_group[25-34] & -0.44 & 0.27 & -0.93 & 0.06 \\
age\_group[35-44] & -0.64 & 0.26 & -1.17 & -0.18 \\
age\_group[45-54] & -0.92 & 0.26 & -1.44 & -0.46 \\
age\_group[55-65] & -0.57 & 0.24 & -1.0 & -0.1 \\
city\_size[10000 - 100000] & 0.29 & 0.18 & -0.02 & 0.63 \\
city\_size[100000+] & 0.07 & 0.2 & -0.33 & 0.41 \\
cnt\_id\_alpha & 0.38 & 0.03 & 0.34 & 0.43 \\
country\_area[Northeast] & -0.4 & 0.23 & -0.8 & 0.06 \\
country\_area[Nortwest] & 0.0 & 0.23 & -0.46 & 0.4 \\
country\_area[South\_Ilands] & -0.32 & 0.2 & -0.69 & 0.06 \\
education[Nodegree] & -0.1 & 0.18 & -0.42 & 0.23 \\
employment[non\_working\_employed] & 0.04 & 0.22 & -0.38 & 0.44 \\
employment[working\_employed] & 0.98 & 0.18 & 0.64 & 1.3 \\
gender[Male] & 0.19 & 0.15 & -0.08 & 0.46 \\
\bottomrule
\end{tabular}

}
\supptable{Results of the Negative Binomial Model for individual contacts on weekday during leisure activities}
\end{table}
\label{tabsi:mon_lei_ind}
\textbf{weekday collective}


\begin{table}[H]
\let\center\empty
\let\endcenter\relax
\centering
\resizebox{.9\width}{!}{\begin{tabular}{lllll}
\toprule
 & mean & sd & hdi\_3\% & hdi\_97\% \\
\midrule
Intercept & 2.46 & 0.26 & 1.99 & 2.97 \\
age\_group[25-34] & -0.52 & 0.25 & -0.99 & -0.04 \\
age\_group[35-44] & -0.66 & 0.24 & -1.1 & -0.2 \\
age\_group[45-54] & -0.82 & 0.25 & -1.3 & -0.39 \\
age\_group[55-65] & -0.55 & 0.23 & -0.96 & -0.13 \\
city\_size[10000 - 100000] & 0.28 & 0.17 & -0.01 & 0.61 \\
city\_size[100000+] & 0.1 & 0.18 & -0.24 & 0.44 \\
cnt\_id\_alpha & 0.39 & 0.02 & 0.35 & 0.44 \\
country\_area[Northeast] & -0.26 & 0.21 & -0.64 & 0.14 \\
country\_area[Nortwest] & 0.12 & 0.21 & -0.26 & 0.53 \\
country\_area[South\_Ilands] & -0.13 & 0.19 & -0.5 & 0.21 \\
education[Nodegree] & -0.06 & 0.16 & -0.35 & 0.25 \\
employment[non\_working\_employed] & 0.09 & 0.2 & -0.25 & 0.48 \\
employment[working\_employed] & 0.99 & 0.16 & 0.7 & 1.31 \\
gender[Male] & 0.14 & 0.13 & -0.11 & 0.38 \\
\bottomrule
\end{tabular}

}
\supptable{Results of the Negative Binomial Model for both individual and collective contacts on weekday during leisure activities}
\label{tabsi:mon_lei_col}
\end{table}
\textbf{weekend individual}


\begin{table}[H]
\let\center\empty
\let\endcenter\relax
\centering
\resizebox{.9\width}{!}{\begin{tabular}{lllll}
\toprule
 & mean & sd & hdi\_3\% & hdi\_97\% \\
\midrule
Intercept & 3.4 & 0.23 & 2.96 & 3.82 \\
age\_group[25-34] & -0.26 & 0.21 & -0.64 & 0.15 \\
age\_group[35-44] & -0.49 & 0.21 & -0.89 & -0.11 \\
age\_group[45-54] & -0.61 & 0.2 & -0.98 & -0.23 \\
age\_group[55-65] & -0.94 & 0.21 & -1.34 & -0.59 \\
city\_size[10000 - 100000] & -0.05 & 0.14 & -0.3 & 0.2 \\
city\_size[100000+] & 0.05 & 0.15 & -0.21 & 0.33 \\
cnt\_id\_alpha & 0.56 & 0.03 & 0.5 & 0.62 \\
country\_area[Northeast] & -0.76 & 0.18 & -1.12 & -0.4 \\
country\_area[Nortwest] & -0.1 & 0.16 & -0.39 & 0.22 \\
country\_area[South\_Ilands] & -0.32 & 0.15 & -0.58 & -0.03 \\
education[Nodegree] & -0.14 & 0.14 & -0.42 & 0.11 \\
employment[non\_working\_employed] & -0.98 & 0.17 & -1.29 & -0.65 \\
employment[working\_employed] & 0.2 & 0.13 & -0.06 & 0.45 \\
gender[Male] & 0.1 & 0.11 & -0.1 & 0.31 \\
\bottomrule
\end{tabular}

}
\supptable{Results of the Negative Binomial Model for individual contacts on weekend during leisure activities}
\label{tabsi:sun_lei_ind}
\end{table}
\textbf{weekend collective}


\begin{table}[H]
\let\center\empty
\let\endcenter\relax
\centering
\resizebox{.9\width}{!}{\begin{tabular}{lllll}
\toprule
 & mean & sd & hdi\_3\% & hdi\_97\% \\
\midrule
Intercept & 3.51 & 0.22 & 3.1 & 3.92 \\
age\_group[25-34] & -0.22 & 0.22 & -0.62 & 0.18 \\
age\_group[35-44] & -0.35 & 0.2 & -0.72 & 0.06 \\
age\_group[45-54] & -0.65 & 0.2 & -1.02 & -0.27 \\
age\_group[55-65] & -0.86 & 0.2 & -1.24 & -0.5 \\
city\_size[10000 - 100000] & 0.09 & 0.13 & -0.15 & 0.36 \\
city\_size[100000+] & 0.11 & 0.15 & -0.16 & 0.4 \\
cnt\_id\_alpha & 0.51 & 0.03 & 0.46 & 0.56 \\
country\_area[Northeast] & -0.37 & 0.18 & -0.71 & -0.05 \\
country\_area[Nortwest] & -0.1 & 0.16 & -0.38 & 0.23 \\
country\_area[South\_Ilands] & -0.22 & 0.15 & -0.53 & 0.03 \\
education[Nodegree] & 0.06 & 0.14 & -0.2 & 0.29 \\
employment[non\_working\_employed] & -0.82 & 0.16 & -1.11 & -0.49 \\
employment[working\_employed] & 0.13 & 0.13 & -0.11 & 0.36 \\
gender[Male] & 0.07 & 0.11 & -0.12 & 0.27 \\
\bottomrule
\end{tabular}

}
\supptable{Results of the Negative Binomial Model for both individual and collective contacts on weekend during leisure activities}
\label{tabsi:sun_lei_col}
\end{table}
\subsubsection*{Transport}
\paragraph*{S1.3}
\label{SI_transport_nb}
\textbf{weekday individual}


\begin{table}[H]
\let\center\empty
\let\endcenter\relax
\centering
\resizebox{.9\width}{!}{\begin{tabular}{lllll}
\toprule
 & mean & sd & hdi\_3\% & hdi\_97\% \\
\midrule
Intercept & 2.24 & 0.42 & 1.42 & 3.03 \\
age\_group[25-34] & -0.95 & 0.37 & -1.63 & -0.26 \\
age\_group[35-44] & -1.36 & 0.38 & -2.07 & -0.67 \\
age\_group[45-54] & -1.35 & 0.36 & -2.0 & -0.67 \\
age\_group[55-65] & -1.42 & 0.35 & -2.02 & -0.72 \\
city\_size[10000 - 100000] & 0.32 & 0.27 & -0.21 & 0.8 \\
city\_size[100000+] & 0.31 & 0.3 & -0.22 & 0.89 \\
cnt\_id\_alpha & 0.27 & 0.02 & 0.23 & 0.32 \\
country\_area[Northeast] & -0.66 & 0.35 & -1.3 & -0.0 \\
country\_area[Nortwest] & 0.33 & 0.33 & -0.27 & 0.96 \\
country\_area[South\_Ilands] & -0.06 & 0.28 & -0.59 & 0.44 \\
education[Nodegree] & -0.15 & 0.25 & -0.62 & 0.32 \\
employment[non\_working\_employed] & -0.28 & 0.32 & -0.85 & 0.32 \\
employment[working\_employed] & 1.42 & 0.28 & 0.85 & 1.89 \\
gender[Male] & 0.57 & 0.24 & 0.13 & 1.02 \\
\bottomrule
\end{tabular}

}
\supptable{Results of the Negative Binomial Model for individual contacts on weekday on public transport}
\label{tabsi:mon_tra_ind}
\end{table}
\textbf{weekday collective}


\begin{table}[H]
\let\center\empty
\let\endcenter\relax
\centering
\resizebox{.9\width}{!}{\begin{tabular}{lllll}
\toprule
 & mean & sd & hdi\_3\% & hdi\_97\% \\
\midrule
Intercept & 2.87 & 0.37 & 2.18 & 3.6 \\
age\_group[25-34] & -1.04 & 0.34 & -1.65 & -0.39 \\
age\_group[35-44] & -1.29 & 0.33 & -1.91 & -0.69 \\
age\_group[45-54] & -1.36 & 0.34 & -1.94 & -0.7 \\
age\_group[55-65] & -1.39 & 0.32 & -1.99 & -0.8 \\
city\_size[10000 - 100000] & 0.12 & 0.24 & -0.35 & 0.53 \\
city\_size[100000+] & 0.14 & 0.27 & -0.36 & 0.64 \\
cnt\_id\_alpha & 0.28 & 0.02 & 0.24 & 0.33 \\
country\_area[Northeast] & -0.68 & 0.3 & -1.3 & -0.15 \\
country\_area[Nortwest] & 0.44 & 0.3 & -0.12 & 0.98 \\
country\_area[South\_Ilands] & -0.13 & 0.25 & -0.58 & 0.35 \\
education[Nodegree] & -0.18 & 0.22 & -0.64 & 0.21 \\
employment[non\_working\_employed] & -0.38 & 0.29 & -0.94 & 0.14 \\
employment[working\_employed] & 1.27 & 0.25 & 0.77 & 1.7 \\
gender[Male] & 0.5 & 0.2 & 0.14 & 0.89 \\
\bottomrule
\end{tabular}

}
\supptable{Results of the Negative Binomial Model for both individual and collective contacts on weekday on public transport}
\label{tabsi:mon_tra_col}
\end{table}

\textbf{weekend individual}


\begin{table}[H]
\let\center\empty
\let\endcenter\relax
\centering
\resizebox{.9\width}{!}{\begin{tabular}{lllll}
\toprule
 & mean & sd & hdi\_3\% & hdi\_97\% \\
\midrule
Intercept & 3.15 & 0.36 & 2.41 & 3.78 \\
age\_group[25-34] & -0.25 & 0.33 & -0.89 & 0.31 \\
age\_group[35-44] & -0.45 & 0.32 & -1.06 & 0.15 \\
age\_group[45-54] & -0.58 & 0.32 & -1.18 & 0.0 \\
age\_group[55-65] & -1.27 & 0.32 & -1.85 & -0.65 \\
city\_size[10000 - 100000] & 0.38 & 0.23 & -0.01 & 0.84 \\
city\_size[100000+] & 0.43 & 0.24 & -0.04 & 0.88 \\
cnt\_id\_alpha & 0.34 & 0.03 & 0.29 & 0.39 \\
country\_area[Northeast] & -0.83 & 0.32 & -1.42 & -0.2 \\
country\_area[Nortwest] & 0.44 & 0.29 & -0.1 & 1.0 \\
country\_area[South\_Ilands] & -0.49 & 0.26 & -1.0 & -0.0 \\
education[Nodegree] & 0.02 & 0.22 & -0.39 & 0.42 \\
employment[non\_working\_employed] & -1.26 & 0.28 & -1.81 & -0.75 \\
employment[working\_employed] & -0.06 & 0.23 & -0.49 & 0.4 \\
gender[Male] & 0.37 & 0.21 & -0.01 & 0.79 \\
\bottomrule
\end{tabular}

}
\supptable{Results of the Negative Binomial Model for individual contacts on weekend on public transport}
\label{tabsi:sun_tra_ind}
\end{table}
\textbf{weekend collective}


\begin{table}[H]
\let\center\empty
\let\endcenter\relax
\centering
\resizebox{.9\width}{!}{\begin{tabular}{lllll}
\toprule
 & mean & sd & hdi\_3\% & hdi\_97\% \\
\midrule
Intercept & 3.2 & 0.34 & 2.54 & 3.8 \\
age\_group[25-34] & -0.43 & 0.32 & -1.02 & 0.17 \\
age\_group[35-44] & -0.62 & 0.31 & -1.21 & -0.04 \\
age\_group[45-54] & -0.52 & 0.3 & -1.08 & 0.08 \\
age\_group[55-65] & -1.16 & 0.3 & -1.75 & -0.64 \\
city\_size[10000 - 100000] & 0.37 & 0.2 & -0.03 & 0.74 \\
city\_size[100000+] & 0.52 & 0.24 & 0.08 & 0.95 \\
cnt\_id\_alpha & 0.36 & 0.03 & 0.31 & 0.4 \\
country\_area[Northeast] & -0.63 & 0.3 & -1.17 & -0.06 \\
country\_area[Nortwest] & 0.47 & 0.27 & -0.06 & 0.95 \\
country\_area[South\_Ilands] & -0.36 & 0.24 & -0.79 & 0.09 \\
education[Nodegree] & 0.09 & 0.21 & -0.3 & 0.48 \\
employment[non\_working\_employed] & -1.22 & 0.27 & -1.71 & -0.72 \\
employment[working\_employed] & -0.01 & 0.23 & -0.43 & 0.43 \\
gender[Male] & 0.37 & 0.19 & 0.02 & 0.72 \\
\bottomrule
\end{tabular}

}
\supptable{Results of the Negative Binomial Model for both individual and collective contacts on weekend on public transport}
\label{tabsi:sun_tra_col}
\end{table}

\subsubsection*{Health activities}
\paragraph*{S1.4}
\label{SI_healt_nb}
\textbf{weekday individual}


\begin{table}[H]
\let\center\empty
\let\endcenter\relax
\centering
\resizebox{.9\width}{!}{\begin{tabular}{lllll}
\toprule
 & mean & sd & hdi\_3\% & hdi\_97\% \\
\midrule
Intercept & 2.49 & 0.4 & 1.7 & 3.21 \\
age\_group[25-34] & -0.28 & 0.36 & -0.91 & 0.46 \\
age\_group[35-44] & -0.97 & 0.36 & -1.65 & -0.32 \\
age\_group[45-54] & -0.73 & 0.34 & -1.37 & -0.07 \\
age\_group[55-65] & -0.88 & 0.33 & -1.47 & -0.26 \\
city\_size[10000 - 100000] & 0.09 & 0.26 & -0.4 & 0.56 \\
city\_size[100000+] & -0.26 & 0.3 & -0.82 & 0.3 \\
cnt\_id\_alpha & 0.28 & 0.02 & 0.24 & 0.32 \\
country\_area[Northeast] & -1.04 & 0.32 & -1.65 & -0.43 \\
country\_area[Nortwest] & 0.16 & 0.32 & -0.44 & 0.77 \\
country\_area[South\_Ilands] & -0.06 & 0.26 & -0.56 & 0.43 \\
education[Nodegree] & -0.32 & 0.23 & -0.77 & 0.12 \\
employment[non\_working\_employed] & -0.34 & 0.29 & -0.92 & 0.17 \\
employment[working\_employed] & 0.92 & 0.26 & 0.46 & 1.41 \\
gender[Male] & 0.67 & 0.2 & 0.29 & 1.03 \\
\bottomrule
\end{tabular}

}
\supptable{Results of the Negative Binomial Model for individual contacts on weekday during health activities}
\label{tabsi:mon_hea_ind}
\end{table}
\textbf{weekday collective}


\begin{table}[H]
\let\center\empty
\let\endcenter\relax
\centering
\resizebox{.9\width}{!}{\begin{tabular}{lllll}
\toprule
 & mean & sd & hdi\_3\% & hdi\_97\% \\
\midrule
Intercept & 2.84 & 0.38 & 2.11 & 3.5 \\
age\_group[25-34] & -0.35 & 0.34 & -0.94 & 0.32 \\
age\_group[35-44] & -0.94 & 0.32 & -1.58 & -0.39 \\
age\_group[45-54] & -0.64 & 0.31 & -1.2 & -0.03 \\
age\_group[55-65] & -0.78 & 0.3 & -1.36 & -0.22 \\
city\_size[10000 - 100000] & -0.01 & 0.22 & -0.42 & 0.38 \\
city\_size[100000+] & -0.39 & 0.26 & -0.84 & 0.09 \\
cnt\_id\_alpha & 0.3 & 0.02 & 0.26 & 0.34 \\
country\_area[Northeast] & -0.9 & 0.29 & -1.44 & -0.37 \\
country\_area[Nortwest] & 0.2 & 0.28 & -0.33 & 0.72 \\
country\_area[South\_Ilands] & -0.03 & 0.24 & -0.49 & 0.42 \\
education[Nodegree] & -0.32 & 0.22 & -0.72 & 0.09 \\
employment[non\_working\_employed] & -0.31 & 0.26 & -0.8 & 0.18 \\
employment[working\_employed] & 0.82 & 0.24 & 0.37 & 1.28 \\
gender[Male] & 0.59 & 0.19 & 0.24 & 0.95 \\
\bottomrule
\end{tabular}

}
\supptable{Results of the Negative Binomial Model for both individual and collective contacts on weekday during health activities}
\label{tabsi:mon_hea_col}
\end{table}

\textbf{weekend individual}


\begin{table}[H]
\let\center\empty
\let\endcenter\relax
\centering
\resizebox{.9\width}{!}{\begin{tabular}{lllll}
\toprule
 & mean & sd & hdi\_3\% & hdi\_97\% \\
\midrule
Intercept & 2.72 & 0.43 & 1.96 & 3.55 \\
age\_group[25-34] & -0.24 & 0.36 & -0.93 & 0.44 \\
age\_group[35-44] & -1.0 & 0.36 & -1.68 & -0.32 \\
age\_group[45-54] & -0.28 & 0.35 & -0.95 & 0.36 \\
age\_group[55-65] & -1.3 & 0.36 & -1.99 & -0.65 \\
city\_size[10000 - 100000] & -0.02 & 0.25 & -0.47 & 0.48 \\
city\_size[100000+] & 0.22 & 0.27 & -0.31 & 0.69 \\
cnt\_id\_alpha & 0.36 & 0.03 & 0.31 & 0.42 \\
country\_area[Northeast] & -1.18 & 0.31 & -1.75 & -0.57 \\
country\_area[Nortwest] & 0.18 & 0.3 & -0.41 & 0.7 \\
country\_area[South\_Ilands] & -0.27 & 0.26 & -0.74 & 0.23 \\
education[Nodegree] & 0.15 & 0.24 & -0.31 & 0.58 \\
employment[non\_working\_employed] & -0.9 & 0.26 & -1.44 & -0.46 \\
employment[working\_employed] & 0.58 & 0.26 & 0.1 & 1.07 \\
gender[Male] & 0.38 & 0.22 & -0.03 & 0.78 \\
\bottomrule
\end{tabular}

}
\supptable{Results of the Negative Binomial Model for individual contacts on weekend during health activities}
\label{tabsi:sun_hea_ind}
\end{table}
\textbf{weekend collective}


\begin{table}[H]
\let\center\empty
\let\endcenter\relax
\centering
\resizebox{.9\width}{!}{\begin{tabular}{lllll}
\toprule
 & mean & sd & hdi\_3\% & hdi\_97\% \\
\midrule
Intercept & 3.23 & 0.39 & 2.47 & 3.94 \\
age\_group[25-34] & -0.47 & 0.34 & -1.1 & 0.18 \\
age\_group[35-44] & -1.05 & 0.33 & -1.74 & -0.48 \\
age\_group[45-54] & -0.44 & 0.33 & -1.04 & 0.18 \\
age\_group[55-65] & -1.11 & 0.32 & -1.76 & -0.54 \\
city\_size[10000 - 100000] & 0.04 & 0.22 & -0.37 & 0.43 \\
city\_size[100000+] & 0.18 & 0.25 & -0.3 & 0.61 \\
cnt\_id\_alpha & 0.39 & 0.03 & 0.33 & 0.45 \\
country\_area[Northeast] & -0.83 & 0.28 & -1.35 & -0.27 \\
country\_area[Nortwest] & 0.22 & 0.28 & -0.3 & 0.74 \\
country\_area[South\_Ilands] & -0.39 & 0.23 & -0.83 & 0.01 \\
education[Nodegree] & -0.08 & 0.21 & -0.45 & 0.32 \\
employment[non\_working\_employed] & -0.93 & 0.24 & -1.44 & -0.52 \\
employment[working\_employed] & 0.36 & 0.23 & -0.06 & 0.79 \\
gender[Male] & 0.33 & 0.19 & -0.06 & 0.65 \\
\bottomrule
\end{tabular}

}
\supptable{Results of the Negative Binomial Model for both individual and collective contacts on weekend during health activities}
\label{tabsi:sun_hea_col}
\end{table}
\subsection{Contact matrices}
\label{SI.2}

To further understand the contact behavior of the participants, and compare it with the pre-pandemic and pandemic period we compute age-stratified contact matrices \cite{jarvis2020quantifying, coletti2020comix}. Participants recorded age information only for individual contacts, for this reason, we focus our analysis only on this subset. Contact matrix elements were computed according to the following equation:

\begin{equation}m_{ij} = \frac{\sum^{T_i}_{t=1}{w_{it}y_{ijt}}}{\sum^{T_i}_{t=1}{w_{it}}}
\end{equation} 

where $y_{ijt}$ denotes the reported number of contacts of the participant $t$ of age $i$ with someone of age $j$. $T_i$ denotes all participants of age $i$, and $w_{it}$ is the post-stratification weight. This accounts for possible under- and over-sampling of age categories or survey day type (either weekday or weekend). 
We also computed contact matrices according to different contact settings: (i) transport, (ii) health facilities, i.e. hospitals or family doctors, (iii) leisure activities, and (iv) essential activities. To compute the contact matrices for the different strata or locations, we considered only the contacts that participants had in a specific location $S$. In particular, the matrix element is given by: 
\begin{equation}m^S_{ij} = \frac{\sum^{T_i}_{t=1}{w_{it}y^S_{ijt}}}{\sum^{T_i}_{t=1}{w_{it}}}\end{equation} 
where $y^S_{ijt}$ is the number of contacts of the participant $t$ of age $i$ with someone of age $j$ in the location $S$. 
To account for sampling variability, we applied a bootstrapping method with replacement to the sample with $n=10000$, and we computed the median of the realizations for each element of the matrix.

\begin{figure}[!h]
  \centering
  \includegraphics[width=1\linewidth]{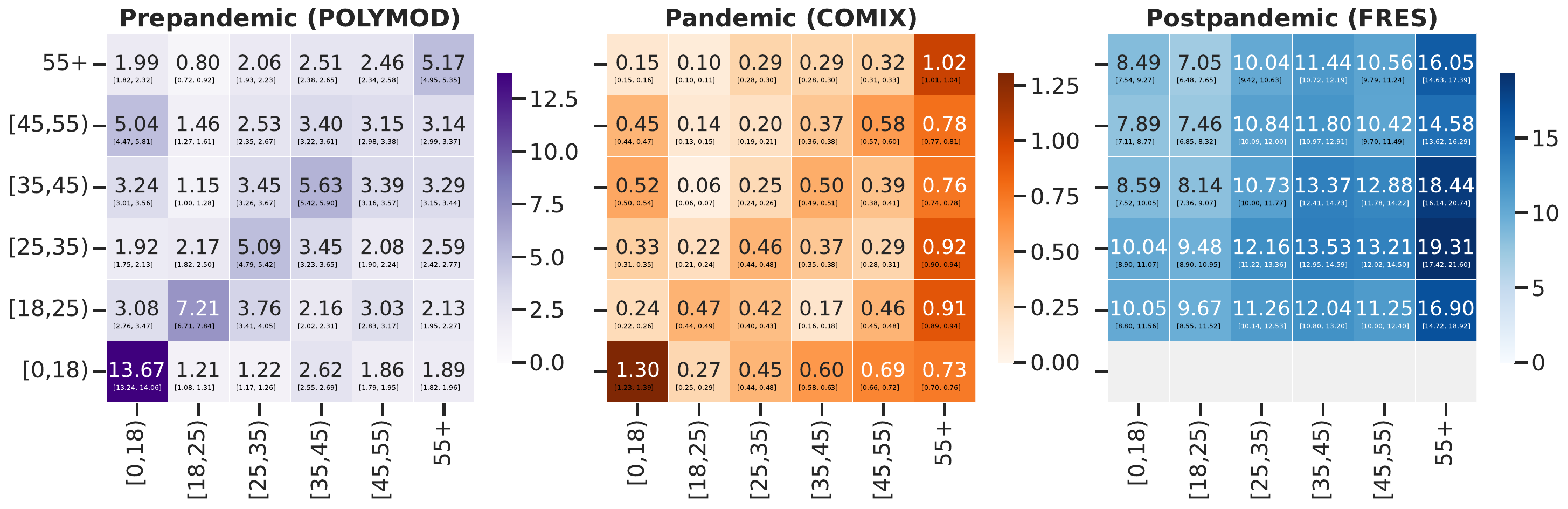}
  \suppfig{Contact patterns before COVID-19 (Prepandemic), during COVID-19 (Pandemic), and after COVID-19 (Post-pandemic). The matrix element ($m_{ij}$) is the median of the bootstrap realizations of the average number of contacts of age group $i$ with participants of age group $j$. For each matrix, we consider only individual contacts over the whole period of the study. The interquartile range is also reported in the parenthesis.}
  \label{fig:cm_comparison_an}
\end{figure}

\begin{figure}[!hb]
  \centering
  \includegraphics[width=1\linewidth]{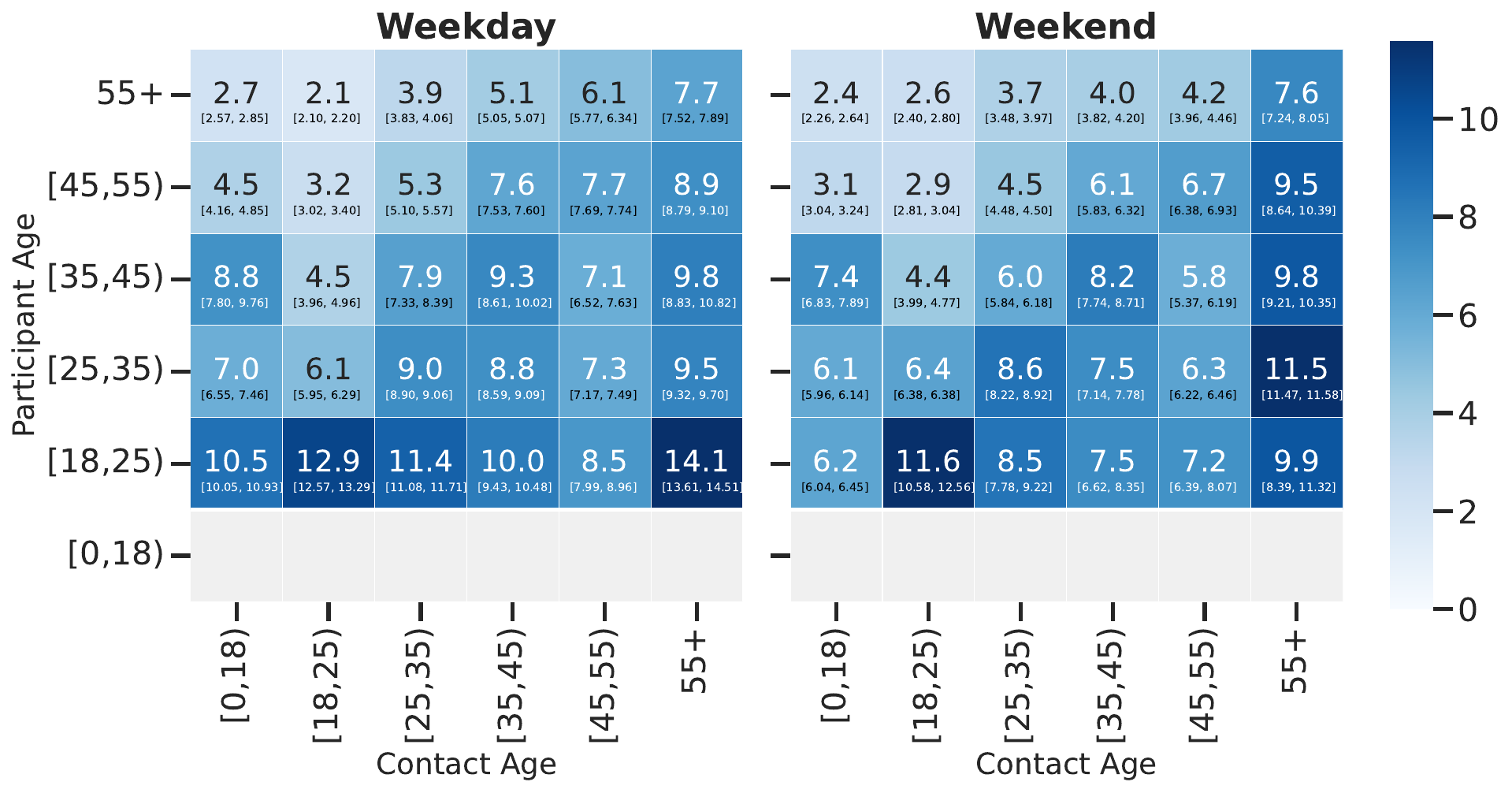}
  \suppfig{Contact matrices on weekday and weekend. The matrix element ($m_{ij}$) is the median of the bootstrap realizations of the average number of contacts of age group $i$ with participants of age group $j$. The interquartile range is also reported in the parenthesis.}
  \label{fig:cm_general_an}
\end{figure}

\paragraph*{SI2.1}
\label{SI2.1}
\begin{figure}[!htb]

  \centering
  \includegraphics[width=1\linewidth]{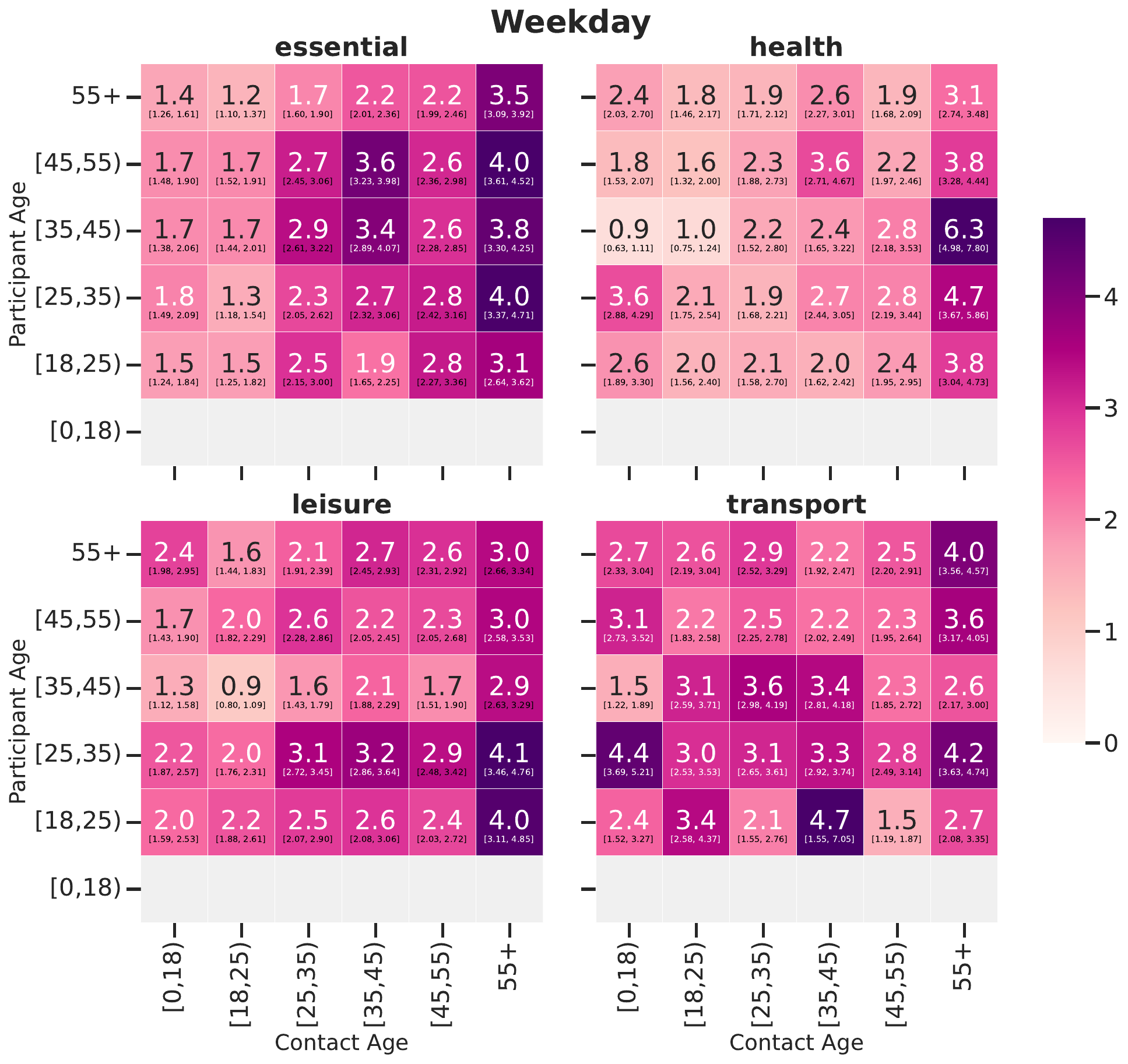}
\suppfig{Contact matrices for different contact locations for weekdays. The matrix element ($m_{ij}$) is the median of the bootstrap realizations of the average number of contacts of age group $i$ with participants of age group $j$. The interquartile range is also reported in the parenthesis.}

\end{figure}

\paragraph*{SI2.2}
\label{SI2.2}
\begin{figure}[!htb]

  \centering
  \includegraphics[width=1\linewidth]{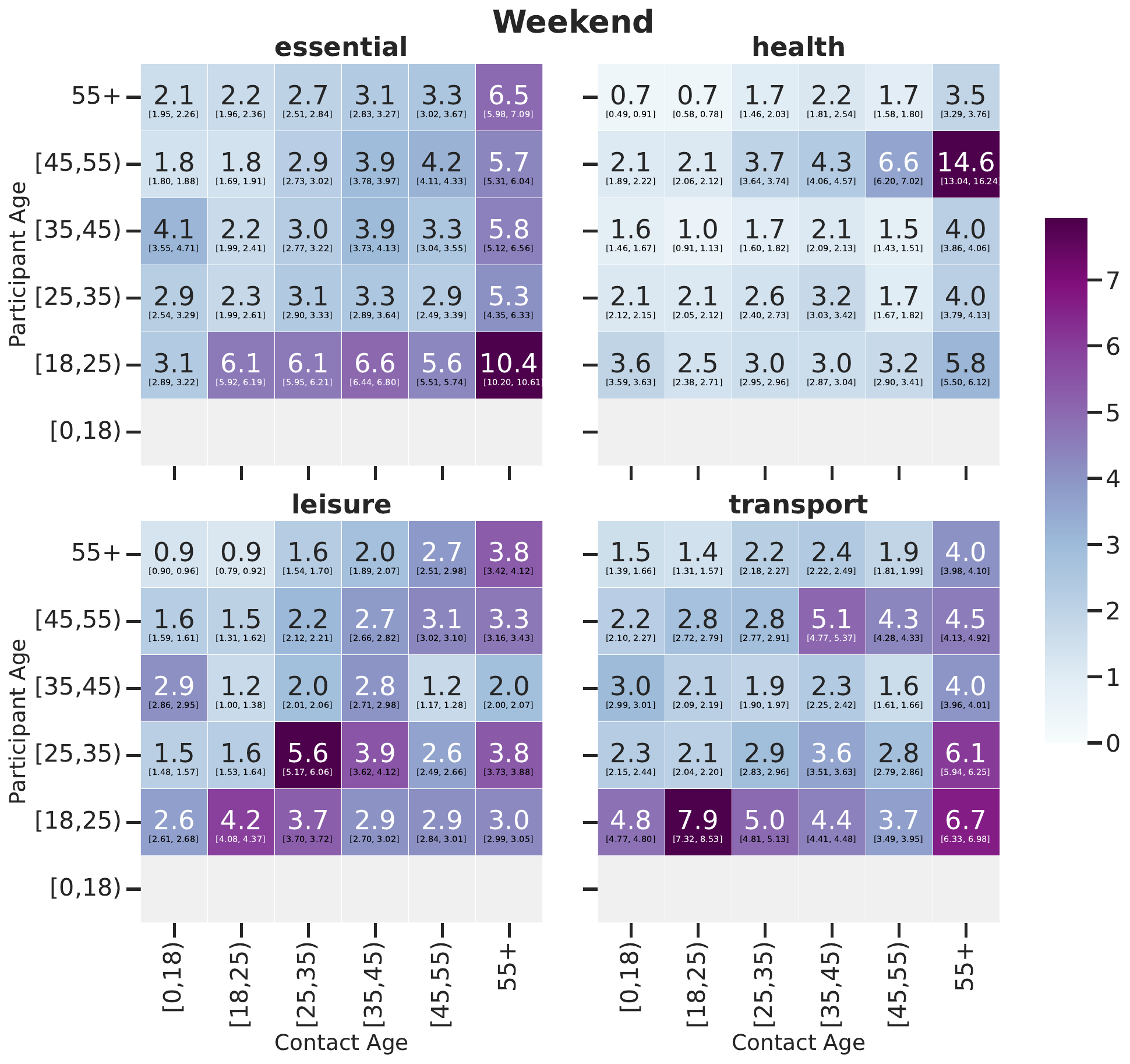}

\suppfig{Contact matrices for different contact locations for weekends. The matrix element ($m_{ij}$) is the median of the bootstrap realizations of the average number of contacts of age group $i$ with participants of age group $j$. The interquartile range is also reported in the parenthesis.}

\end{figure}

\clearpage
\bibliography{FRES2024}

\end{document}